\documentclass[fleqn,usenatbib]{mnras}
\usepackage[T1]{fontenc}
\usepackage{ae,aecompl}

\usepackage[utf8]{inputenc}
\usepackage{physics}
\usepackage{verbatim}
\usepackage{tikz}
\usepackage{amssymb}
\usepackage{array}
\usepackage{siunitx}
\usepackage{etoolbox}
\usepackage{threeparttable}
\usepackage{booktabs}
\usepackage{multicol}
\usetikzlibrary{backgrounds}

\makeatletter
\patchcmd{\NAT@citex}
  {\@citea\NAT@hyper@{%
     \NAT@nmfmt{\NAT@nm}%
     \hyper@natlinkbreak{\NAT@aysep\NAT@spacechar}{\@citeb\@extra@b@citeb}%
     \NAT@date}}
  {\@citea\NAT@nmfmt{\NAT@nm}%
   \NAT@aysep\NAT@spacechar\NAT@hyper@{\NAT@date}}{}{}
   
\patchcmd{\NAT@citex}
  {\@citea\NAT@hyper@{%
     \NAT@nmfmt{\NAT@nm}%
     \hyper@natlinkbreak{\NAT@spacechar\NAT@@open\if*#1*\else#1\NAT@spacechar\fi}%
       {\@citeb\@extra@b@citeb}%
     \NAT@date}}
  {\@citea\NAT@nmfmt{\NAT@nm}%
   \NAT@spacechar\NAT@@open\if*#1*\else#1\NAT@spacechar\fi\NAT@hyper@{\NAT@date}}
  {}{}
\makeatother

\let\oldciteauthor=\citeauthor
\def\newciteauthor#1{\hypersetup{citecolor=black}\oldciteauthor{#1}}
\def\citeauthor#1{\newciteauthor{#1}\hypersetup{citecolor=blue}}

\newcommand{\e}[1]{\times 10^{#1}}
\newcommand{\pfrac}[2]{\left(\frac{#1}{#2}\right)}
\newcommand{\pfracp}[3]{\left(\frac{#1}{#2}\right)^{#3}}
\newcommand{\mbh}{M_{\rm BH}}
\newcommand{\mdbh}{\dot{M}_{\rm BH}}

\newcommand{\edd}{\text{Edd}}
\newcommand{\mdedd}{\dot{M}_\edd}

\newcommand{\xin}{\text{in}}
\newcommand{\xout}{\text{out}}

\newcommand{\msun}{M_\odot}

\newcommand{\email}[1]{\mbox{\href{mailto:#1}{#1}}}

\newcolumntype{C}{>{$}c<{$}} 

\title[Tidal disruptions from MBHBs]{Tidal disruption events from massive black hole binaries: predictions for ongoing and future surveys}
\author[S.\ Thorp, E.\ Chadwick and A.\ Sesana]{Stephen Thorp$^{1,2}$\thanks{E-mail: \email{sjt202@cam.ac.uk}}, Eli Chadwick,$^{1,3}$ and Alberto Sesana$^{1,4}$\\
$^1$School of Physics \& Astronomy, University of Birmingham, Birmingham, B15 2TT, UK\\
$^2$Institute of Astronomy, University of Cambridge, Madingley Road, Cambridge, CB3 0HA, UK\\
  $^3$Scientific Computing Department, Science and Technology Facilities Council, Rutherford Appleton Laboratory, Didcot, OX11 0QX, UK\\
  $^4$Dipartimento di Fisica ``G. Occhialini'', Universit\`a degli Studi di Milano-Bicocca, Piazza della Scienza 3, 20126 Milano, Italy}

\date{Accepted \dots Received \dots; in original form \dots}

\pubyear{2018}

\begin{document}
\label{firstpage}
\pagerange{\pageref{firstpage}--\pageref{lastpage}}
\maketitle

\begin{abstract}
  We compute the expected cosmic rates of tidal disruption events induced by individual massive black holes (MBHs) and by MBH binaries (MBHBs) -- with a specific focus on the latter class -- to explore the potential of TDEs to probe the cosmic population of sub-pc MBHBs. Rates are computed by combining MBH and MBHB population models derived from large cosmological simulations with estimates of the induced TDE rates for each class of objects. We construct empirical TDE spectra that fit a large number of observations in the optical, UV and X-ray and consider their observability by current and future survey instruments. Consistent with results in the literature, and depending on the detailed assumption of the model, we find that \textit{LSST} and {\it Gaia} in optical and \textit{eROSITA} in X-ray will observe a total of $3000$--$6000$, $80$--$180$ and $600$--$900$ TDEs per year, respectively. Depending on the survey, one to several percent of these are prompted by MBHBs. In particular both \textit{LSST} and \textit{eROSITA} are expected to see 150--450 MBHB induced TDEs in their respective mission lifetimes, including 5--100 repeated flares. The latter provide an observational sample of binary candidates with relatively low contamination and have the potential of unveiling the sub-pc population of MBHBs in the mass range $10^5\msun<M<10^7\msun$, thus informing future low frequency gravitational wave observatories.    
\end{abstract}


\section{Introduction}
\label{sec:introduction}
Tidal disruption events (TDEs) occur when the tidal forces exerted on a star upon close approach to a massive black hole (MBH) overwhelm its self gravity and pull it apart.  First predicted by \citet{hills75}, a steady stream of theoretical work followed throughout the 1970s and 1980s with \citet{rees88,rees90} and \citet{phinney89} notably predicting a characteristic visible flare which decays as $t^{-5/3}$. The first observations came in the late 1990s, with the \textit{ROSAT} X-ray telescope discovering four candidates in this era \citep{bade96,komossabade99,komossagreiner99,grupethomas99,greinerschwarz00}. Optical and UV detections have followed more recently \citep[see][for a review of observational progress]{komossa15}, yielding a few tens of candidate flares across the electromagnetic spectrum. \citet{auchettl17} present a larger sample of $\sim70$ X-ray events which have at one point been identified as TDEs, but find that only a fraction of these can be confidently classed as TDEs\footnote{An evolving list of events which have at one point been identified as TDEs can be found at The Open TDE Catalog (\url{https://tde.space/}).}.
\par
Of the observed events, one -- SDSS J120136.02+300305.5 \citep{saxton12} -- has been proposed by \citet{2014ApJ...786..103L} as having a massive black hole binary (MBHB) origin, on the basis of dips in the light curve, a phenomenon suggested by \citet{liu09} as a way of distinguishing a binary-induced disruption. Light curve interruption is given further consideration by \citet{vigneron18}, with particular focus on the effect of binary orbital separation. More recently, the transient ASASSN-15lh has been suggested as arising from disruption by an MBHB \citep{coughlinarmitage18}, but numerous alternative scenarios exist for this event \citep{dong16,leloudas16,coughlin18}.
\par
Whilst a single MBH sitting at the centre of a galaxy disrupts stars at a rate of $10^{-5}$--$10^{-4}$~yr$^{-1}$ \cite[e.g.][]{magorrian99,donley02,wang04,gezari08,stone16}, when an MBHB is present, it has been shown that the tidal disruption rate can be increased by orders of magnitude by three body interactions. The physics of the process was first investigated by \cite{2005MNRAS.358.1361I} and later further explored by several authors employing either three body scattering \citep{chen09,chen11,weggbode11} or N-body simulations \citep{2017ApJ...834..195L}. \cite{2013ApJ...767...18L} provided a comprehensive description of the behaviour of the tidal disruption rate during galaxy mergers. As the two MBHs spiral in through dynamical friction, TDEs become more common, due to the perturbation of the stellar distribution that arises in the merger process. This phase is followed by a burst of TDEs, occurring when the two MBHs form a bound binary and start scouring the central cusp of bound stars. When the cusp has been erased and the loss cone is depleted, the now hard binary disrupts stars at a much lower rate \citep{2008ApJ...676...54C}, although possibly comparable or higher than the typical rate for single MBHs \citep{coughlin18,darbha18}.  
\par
In particular, \citet{chen11} parametrize the tidal disruption rates arising from pairing binaries as a function of several key system properties, including binary mass, mass ratio and the profile of the surrounding stellar distribution. Compared to isolated MBHs, sub-pc binaries can produce, for a relatively short period of time (up to few Myr), a TDE rate enhancement of a few orders of magnitude, up to $\sim0.1$~yr$^{-1}$ when the most favourable conditions are met. Put into a cosmological perspective, these results imply that a non-negligible fraction of TDEs might be due to MBHBs or, reversing the argument, that TDEs might be preferential signposts for identifying sub-pc MBHBs. Moreover, an MBHB could trigger multiple disruptions over several years, something which would be extremely unlikely to occur in a galaxy containing a single MBH \citep[see also][]{weggbode11}. Such a recurrent tidal flare from a galaxy could provide strong evidence for the presence of an MBHB in its centre, and may allow us to study the MBHB's parameters. However, several alternative mechanisms for the triggering of multiple disruptions exist -- e.g., the capture and successive disruption of both members of a stellar binary \citep{mandel15,wu18}, the stabilization of an eccentric disc of stars in a galactic nucleus \citep[which can lead to temporarily enhanced TDE rates, as discussed in][]{madigan18}, or the instantaneous loss cone refill expected due to the recoil which follows an MBHB merger \citep{stone11}.
\par
We aim in this study to employ these results in estimating the number of TDEs detectable by \textit{LSST} and \textit{eROSITA}, with a particular emphasis on those arising from MBHBs. With \textit{LSST} covering the optical sky, and \textit{eROSITA} covering the X-ray, scope exists for a large-scale multi-band investigation of tidal disruptions. The possibility of applying these instruments in the search for TDEs has been given prior consideration, promising thousands of detections yearly with \textit{LSST}, and hundreds with \textit{eROSITA} \citep[e.g.][]{vanvelzen11,khabibullin14,mageshwaranmangalam,mangalam18}. However only \citet{weggbode11} have previously considered the search for MBHB induced disruptions, predicting the detection of only a few of these events per year with \textit{LSST}.
\par
In Section \ref{sec:tdrates}, we summarize the key results which we use to calculate the tidal disruption rate for a particular MBH or MBHB. In Section \ref{sec:populations}, we extract populations of galactic centre MBHs and MBHBs from one of the Millennium--II galaxy catalogues, and use this to build a distribution of TDE rates. Section \ref{sec:emission} describes our phenomenological approach to the modelling of TDE emission across the electromagnetic spectrum, with Section \ref{sec:detectors} describing our treatment of the different surveys. Finally, Section \ref{sec:results} presents the detection rates we predict for each of the surveys, and considers (in Section \ref{sec:distinction}) the possibility of recurrent flares being used to discriminate between MBHs and MBHBs.

\section{TDE Rates}
\label{sec:tdrates}
The tidal disruption radius of a black hole (i.e. the distance from it at which a star will be disrupted) is given by
\begin{align}
    r_t&=r_*\pfracp{M_{\rm BH}}{m_*}{1/3} \nonumber \\
    &\approx 5\e{-6}\pfrac{r_*}{R_\odot}\pfracp{m_*}{\msun}{-1/3}\pfracp{M_{\rm BH}}{10^6M_\odot}{1/3}~\si{pc},
    \label{eq:rt}
\end{align}
where $M_{\text{BH}}$ is the black hole mass, $r_*$ and $m_*$ are the stellar radius and mass, and $R_\odot$ and $\msun$ are the solar radius and mass \citep{rees88}.
In order to calculate the rate at which a single black hole of a particular mass, $M_{\text{BH}}$, disrupts stars, we employ the empirically derived results of \citet{stone16}, who find a rate of
\begin{equation}
	\dot{\mathcal{N}}_{s} = 1.86\times10^{-4}\,\left(\frac{M_{\rm BH}}{10^6M_\odot}\right)^{-0.404}~\text{yr}^{-1},
	\label{eq:ntdsingle}
\end{equation}
where we use $\mathcal{N}_s$ to represent the number of stars disrupted by a particular MBH. This is derived based on observations which cover galaxies containing black holes of $M_{\rm BH}\geq10^6M_\odot$, so should be treated with caution below this limit. More generally, we also adopt an upper mass limit of $M_\text{crit}=6\times10^7M_\odot$, above which a (non-rotating) black hole would be unable to support an accretion disc following a disruption, and the star is considered to be swallowed without producing a TDE. Since candidate TDEs from black holes with masses down to around $10^5M_\odot$ exist, we assume this as our lower mass limit. In order to safely extrapolate the TDE rate of \citet{stone16} below $10^6M_\odot$, we follow \citet{babak17} in assuming that a black hole cannot grow by more than an $e$-fold over a Hubble time, $t_H=13.4\times10^9$~yr \citep[with this value based on the cosmology used in][]{guo11}. As such, for lower mass MBHs, where the scaling of equation (\ref{eq:ntdsingle}) would cause them to grow by more than $M_{\rm BH}/e$ over a Hubble time, we cap the TDE rate to prevent this. In doing so, we must make an assumption about the fraction of the stellar mass which is absorbed by the MBH during a disruption. Assuming a typical star of $1M_\odot$, we adopt a default model in which $0.45M_\odot$ is absorbed per disruption, with this being based on the assumption that 10\% of the $\sim0.5M_\odot$ of bound material is lost due to winds \citep[e.g.][]{strubbe09,lodatorossi11}. This gives an effective TDE rate of
\begin{multline}
	\dot{\mathcal{N}}_{s} = \min\left[\frac{(M_{\rm BH}/M_\odot)\times e^{-1}}{0.45\times13.4\times10^9},\right.\\\left.\vphantom{\frac{(M_{\rm BH}/M_\odot)\times e^{-1}}{0.45\times13.4\times10^9}}1.86\times10^{-4}\,\left(\frac{M_{\rm BH}}{10^6M_\odot}\right)^{-0.404}\right]~\text{yr}^{-1}.
	\label{eq:ntdsinglee}
\end{multline}
The $0.45M_\odot$ of growth per disruption assumed above is likely an upper limit for solar-type stars. As a counterpoint, we also repeat some of our calculations for a lower limit of $0.15M_\odot$. This is based on the loss of 70\% of bound material due to winds \citep[e.g.][]{lodatorossi11}.
\par

Assuming a constant fraction of material lost, independent of the TDE properties, is certainly an oversimplification. It is expected that the the amount of material expelled by winds strongly depends on the mass fallback rate onto the black hole following a disruption \citep{dotanshaviv11,lodatorossi11}. This, in turn, depends on black hole mass, with \citet{2018MNRAS.478.3016W} finding that lower mass black holes could experience a sustained phase of highly super-Eddington accretion, whilst more massive black holes with $M_{\text{BH}}\gtrsim10^7M_\odot$ are unlikely to experience super-Eddington accretion at all. As such, we concede that it would be more physically accurate to adopt a scaling where the fraction of mass lost depends on $M_{\text{BH}}$, with a higher mass corresponding to lower fallback rate, a weaker wind, and consequently less mass lost. Given the current state of knowledge of super-Eddington accretion flows and winds, any description of this dependence would require some ad-hoc assumption for the fallback rate and wind loss scaling with $M_{\text{BH}}$. A model for the former could be obtained from \citet{lodatorossi11} or by interpolating fig. 1 of \citet{2018MNRAS.478.3016W}. For the latter, one could use the fitting formula given by \citet[eq.~28]{lodatorossi11} derived from the results of \citet{dotanshaviv11}. We note, however, that any such scaling for the wind loss would predict nearly 100\%  mass loss for $M_{\text{BH}}\lesssim10^6M_\odot$, meaning that applying it in equation (\ref{eq:ntdsinglee}) would leave equation (\ref{eq:ntdsingle}) completely uncapped at masses for which it is just an extrapolation of the results of \citet{stone16}. This would severely inflate the TDE rate at the low end of the MBH mass function, which is generally inconsistent with the fact that most observed TDEs can be attributed to MBHs with $M>10^6\msun$. Besides this, the main focus of the paper is investigating TDEs from MBHBs. Those are intrinsically dominated by high mass systems (see, Section \ref{sec:populations-binarytdrate} and figure \ref{fig:massfuncs}) and their statistics are unaffected by the behaviour of the low mass end of the single MBH TDE rate. In light of these considerations, we keep the constant mass loss fraction as an acceptable working hypothesis.

\par
When considering disruptions arising from binaries, we employ the theory developed by \citet{chen11} for sub-pc, pairing systems. They assume that an MBHB is surrounded by a stellar cusp which follows a density profile, $\rho(r)\propto r^{-\gamma}$, within the binary's sphere of influence. The $\gamma$ parameter introduced here is one of the key variables defining the tidal disruption rate for a particular binary, along with the primary black hole mass, $M_1$, and mass ratio, $q=M_2/M_1$. By modelling the decay of the binary self-consistently during the cusp erosion, they find that an approximately constant TDE rate is sustained for a period of time that depends on the aforementioned parameters, followed by a steep decay. Using their parametrisation, we therefore adopt a disruption rate described over time by a step-function of amplitude
\begin{equation}
  \dot{\mathcal{N}}_{b} \approx 6\times10^{-6}\,(3-\gamma)^{-2}q^{\frac{4-2\gamma}{3-\gamma}}\left(\frac{M_1}{M_\odot}\right)^{2/3}~\text{yr}^{-1},
  \label{eq:tdratechen}
\end{equation}
and duration 
\begin{equation}
	t_D = 3043\,(3-\gamma)^{3/2}(1+q)^{-1/2}q^{\frac{3\gamma-6}{6-2\gamma}}\left(\frac{M_1}{M_\odot}\right)^{1/4}~\text{yr}.
	\label{eq:tdutycycle}
\end{equation}
for a total number of disrupted stars of
\begin{equation}
	\mathcal{N}_{b} \approx 0.02\,(3-\gamma)^{-1/2}q^{\frac{2-\gamma}{6-2\gamma}}\left(\frac{M_1}{M_{\odot}}\right)^{11/12},
	\label{eq:ntd}
\end{equation}
resulting from the combination of equations \eqref{eq:tdratechen} and \eqref{eq:tdutycycle}. Numbers obtained with the simplified equation \eqref{eq:ntd} are consistent with the integrated numbers of TDEs found by \citet{chen11} when consistently evolving the MBHBs within the stellar cusp. 

We stress that \citet{chen11} only model the cusp erosion phase, which leads to significant binary shrinkage, although this is generally insufficient to prompt coalescence via GW emission \citep[see, e.g.,][]{2010ApJ...719..851S}. The binary will then continue to harden via scattering of unbound stars and TDEs will still occur, albeit at a lower rate. The contribution of this later phase to the TDE rate from MBHBs is discussed in Appendix \ref{app:nonenhancedbinaries}. Unless otherwise stated, rates derived throughout the paper only account for the erosion phase of the bound cusp by the newly formed sub-pc MBHB.

\begin{figure*}
\includegraphics[width=0.32\linewidth]{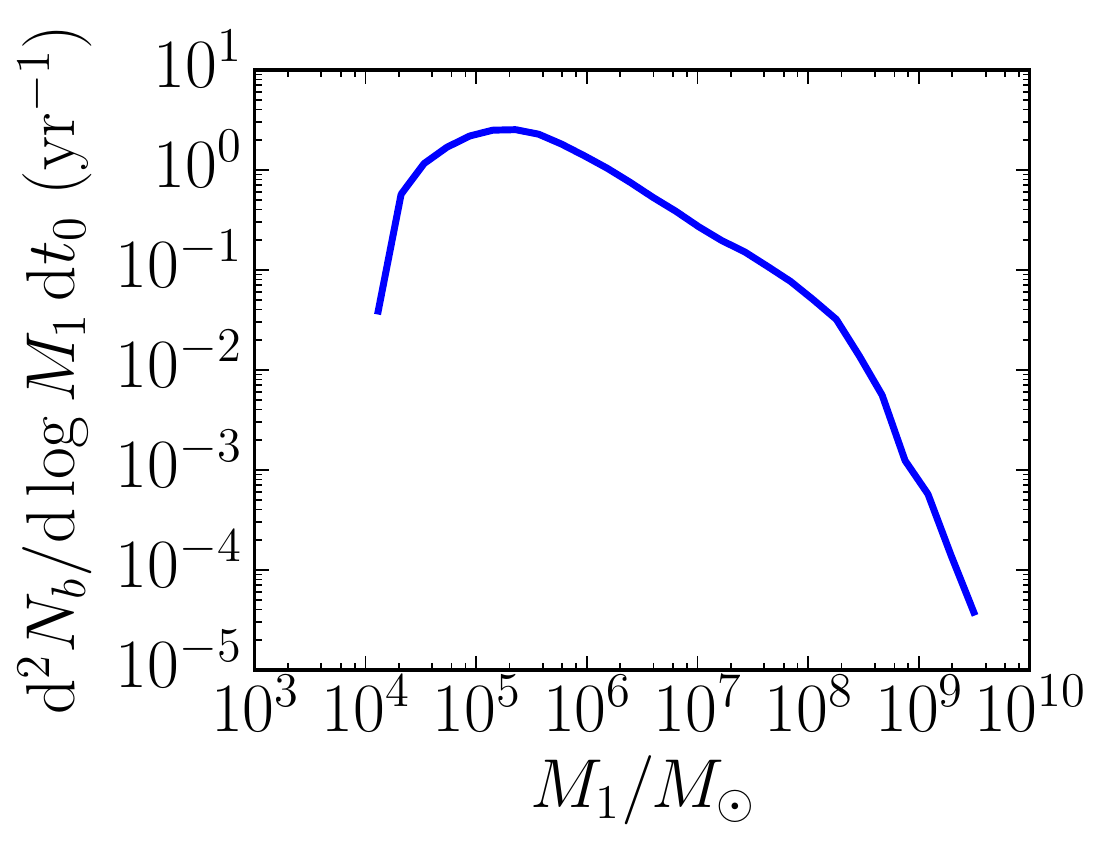}
\includegraphics[width=0.32\linewidth]{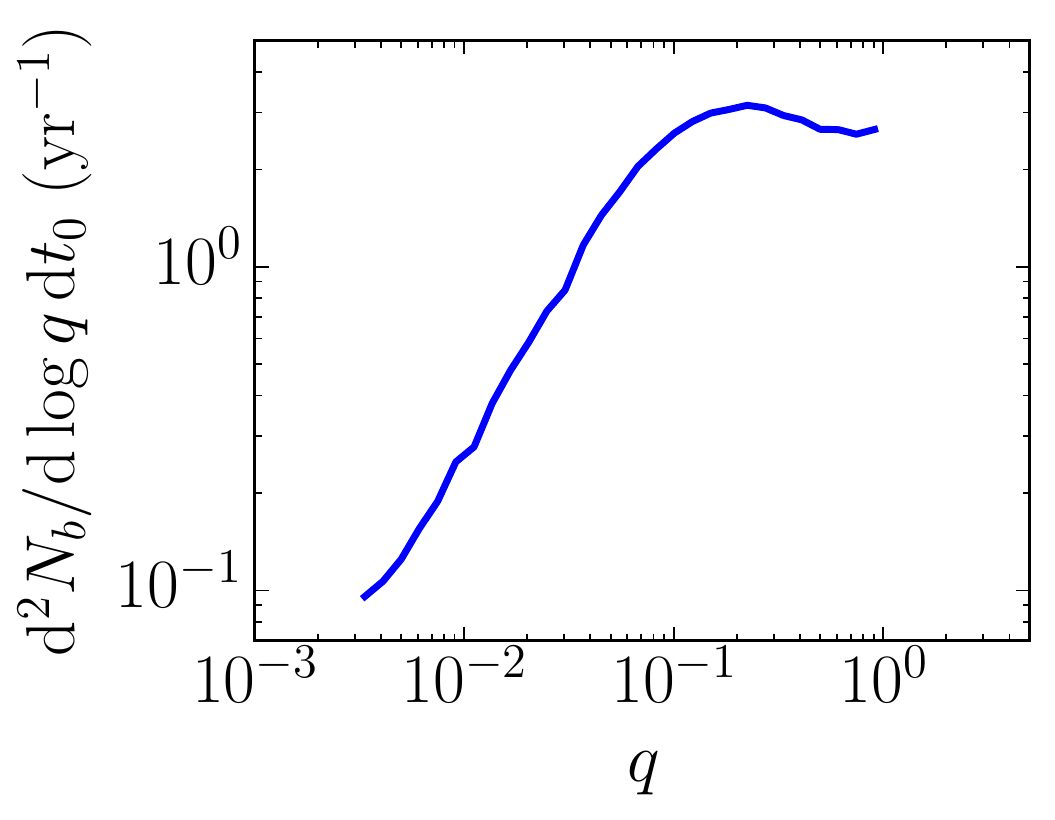}
\includegraphics[width=0.32\linewidth]{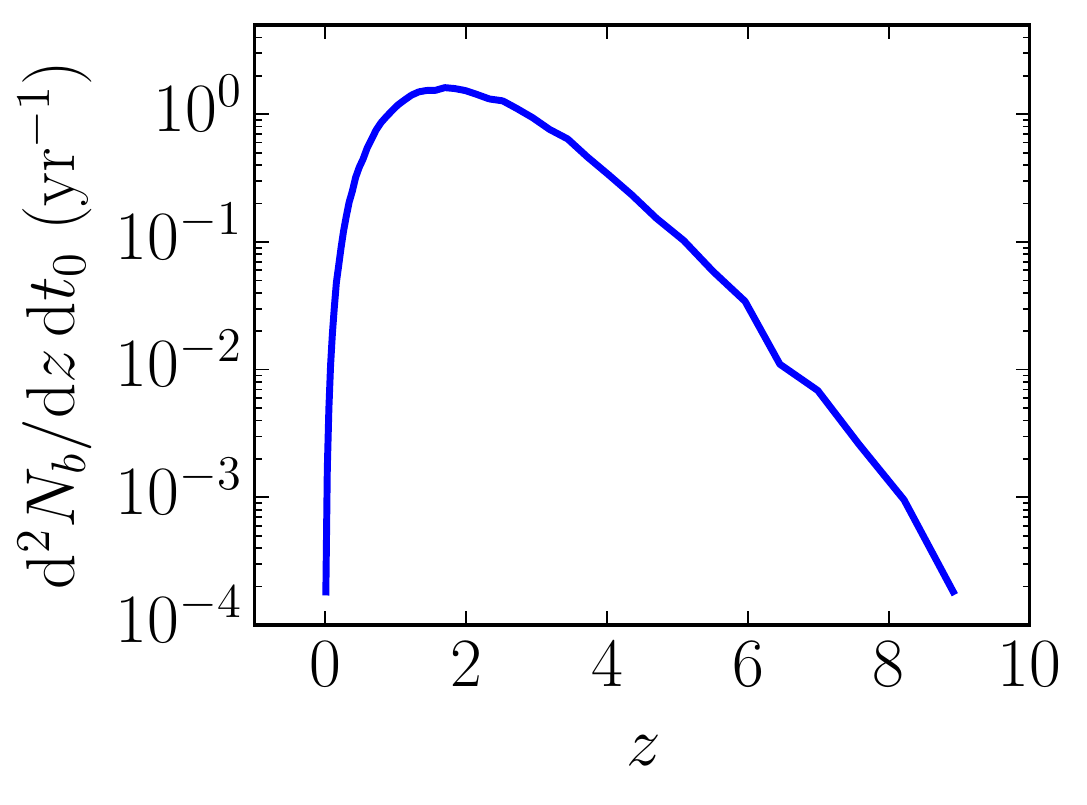}
\caption{Differential MBHB merger rate as a function of primary black hole mass (left), mass ratio (centre), and redshift (right) in the observer frame.}
\label{fig:dNdlogBH}
\end{figure*}

\section{MBH/MBHB Populations}
\label{sec:populations}
To model the population of MBHs and MBHBs along the cosmic history, we employ the \texttt{guo2010a} catalogue of the Millennium--II simulation \citep{millennium06,millenniumii09,guo11}. This has sufficient resolution to distinguish galaxies centred on black holes with masses down to a few $\times 10^4\msun$. Although recent progresses in high performance computing made possible to perform fully hydrodynamical simulations on cosmological comoving volumes of $\sim$100~Mpc a side \citep[e.g.][]{2014MNRAS.444.1518V,2015MNRAS.446..521S}, due to resolution limitation, those simulations start to be incomplete below galaxy masses of $\sim 10^{10}\msun$, corresponding to central MBH masses of $\sim 10^6\msun$. Therefore, any MBHB mass function extracted from them would be incomplete at low masses, especially for very unequal mass ratios (for example, many $10^6\msun-10^5\msun$ binaries are likely to be missing due to the limiting resolution preventing the formation of the typical host galaxy of a $10^5\msun$ MBH). Conversely, the Millennium--II is essentially complete down to much lower masses, allowing a reasonable reconstruction of the MBHB mass function down to $M_{\text{BH}}\sim 10^5\msun$ (see figure \ref{fig:dNdlogBH}), in the mass range relevant for this study. We also stress that we do not expect incompleteness at $M\lesssim10^5\msun$ to be an issue in our calculation. Given the mass scaling of the number of MBHB TDEs in equation \eqref{eq:ntd}, any low mass black holes lost due to resolution will be extremely sub-dominant in their contribution to the overall TDE cosmic population. Throughout this work, we adopt the same cosmology as \citet{guo11}, i.e.\ $h = H_0/(100~\text{km\,s}^{-1}\,\text{Mpc}^{-1})=0.73$, $\Omega_\Lambda=0.75$ and $\Omega_M=0.25$\footnote{Although this cosmology is quite outdated, the precise values of cosmological parameters will affect our results at the level of a few percent at most.}.

\subsection{MBHB Merger Rate}

We extract data on galaxy merger events from this simulation; after filtering to remove insignificant events, we are left with $N_b=169,435$ mergers across the history of the Millennium--II simulation. Since the \texttt{guo2010a} model already discriminates mergers leading to a satellite disruption from those that are efficiently completed, we assume that each of these events forms an MBHB that reaches sub-pc separation and interacts efficiently with the central distribution of stars. We caution, however, that this assumption might be optimistic, since MBHs delivered by very minor mergers might effectively stall at kpc separations instead of reaching the centre of the merger remnant \citep{2011ApJ...729...85C,2018MNRAS.475.4967T}. 

The properties $M_1$ (primary mass, in solar masses), $q$ (mass ratio), and $z$ (redshift) are used to place each merger in a 3D grid. We bin $M_1$ and $q$ in 30 logarithmically spaced bins, with $M_1 \in [10^4,10^{10}] \msun$ and $q\in[0.003,1]$. The redshift, $z$, is binned according to the redshift of each Millennium--II snapshot.

Dividing the number of events in each bin by the comoving volume, $V_c$, of the Millennium--II simulation gives the approximate comoving merger density per unit mass, mass ratio and redshift:
\begin{equation}
    \frac{\dd[4]{N_b}}{\dd{\log M_1}\,\dd{\log q}\,\dd{z}\,\dd{V_c}}.
\end{equation}

For observational purposes, we want to know how many mergers are observed on Earth per unit time. To obtain this quantity, we use standard cosmological functions \citep{hogg99} to obtain:
    \begin{multline}
        \frac{\dd[4]{N_b}}{\dd{\log M_1}\,\dd{\log q}\,\dd{z}\,\dd{t_0}}\\= \frac{\dd[4]{N_b}}{\dd{\log M_1}\,\dd{\log q}\,\dd{z}\,\dd{V_c}}\cdot\dv{V_c}{z}\cdot\dv{z}{t_r}\cdot\dv{t_r}{t_0} ,
        \label{eq:d4Ndt0}
    \end{multline}
where $t_0$ is the time in Earth's rest frame, and $t_r=t_0/(1+z)$ is the time in the rest frame of the source.

To visualize this distribution, we can integrate out two of the three parameters $M_1$, $q$ and $z$ to get one-dimensional distributions. These are shown in Fig.\ \ref{fig:dNdlogBH}.

The mass function of merging binaries follows a typical Schechter function with a characteristic mass of about $10^8M_\odot$. The drop-off below $M\approx 2\times10^{5} M_\odot$ is due to the resolution limit of the simulation, as described above. 

The mass ratio distribution shows a plateau in the range $q\in[0.1,1]$, with a steep drop at lower $q$, consistent with the fact that the more frequent minor mergers often lead to the disruption of the satellite before the merger is complete. Although one could imagine that the MBH carried by the secondary galaxy could still spiral in and form a binary, progressive mass stripping has been found to severely affect the efficiency of dynamical friction, effectively stalling the orbital decay process and ending with the deposition of the MBH far from the nucleus of the merger remnant \citep{2018MNRAS.475.4967T}. We further notice that, even if the MBH from the secondary galaxy is able to sink efficiently through the primary and form a binary, it is likely that the black hole mass ratio resulting from such a minor merger would be sufficiently low that relativistic precession in the cusp would counter any TDE rate enhancement due to the Kozai effect \citep[$q\lesssim 0.01$ is found to lead to near total suppression by][]{chen11}.

The redshift distribution peaks around $z=2$, tracing the evolution of the comoving volume shell with redshift. 

Integrating out all three of $M_1$, $q$ and $z$ from equation \eqref{eq:d4Ndt0}  gives the global merger rate as observed on Earth:
\begin{equation}
    \dv{N_b}{t_0}= \iiint\frac{\dd[4]{N_b}}{\dd{\log M_1}\,\dd{\log q}\,\dd{z}\,\dd{t_0}}\dd{\log M_1}\,\dd{\log q}\,\dd{z}.
    \label{eq:dNdt0}
\end{equation}
For this specific MBHB population model, we get $\dd{N_b}/\dd{t_0}\approx4~\si{yr^{-1}}$.

\subsection{MBHB TDE Rate}
\label{sec:populations-binarytdrate}

\begin{figure*}
\includegraphics[width=0.32\linewidth]{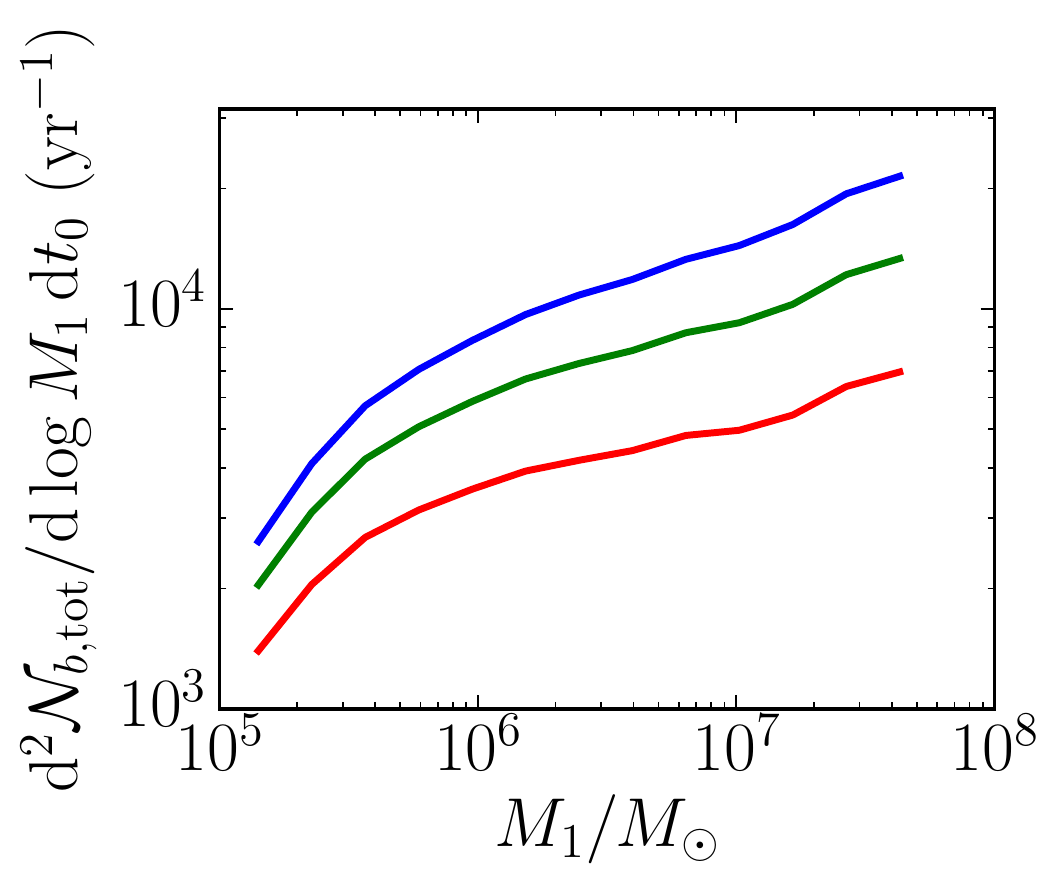}
\includegraphics[width=0.32\linewidth]{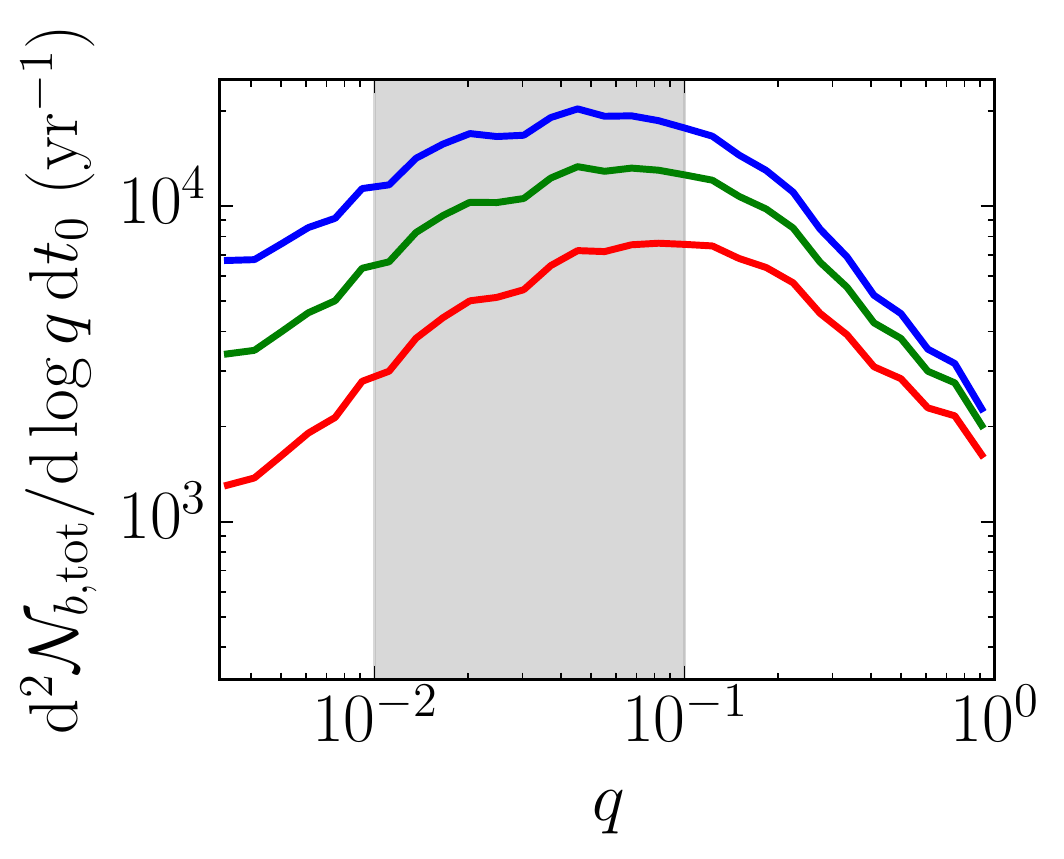}
\includegraphics[width=0.32\linewidth]{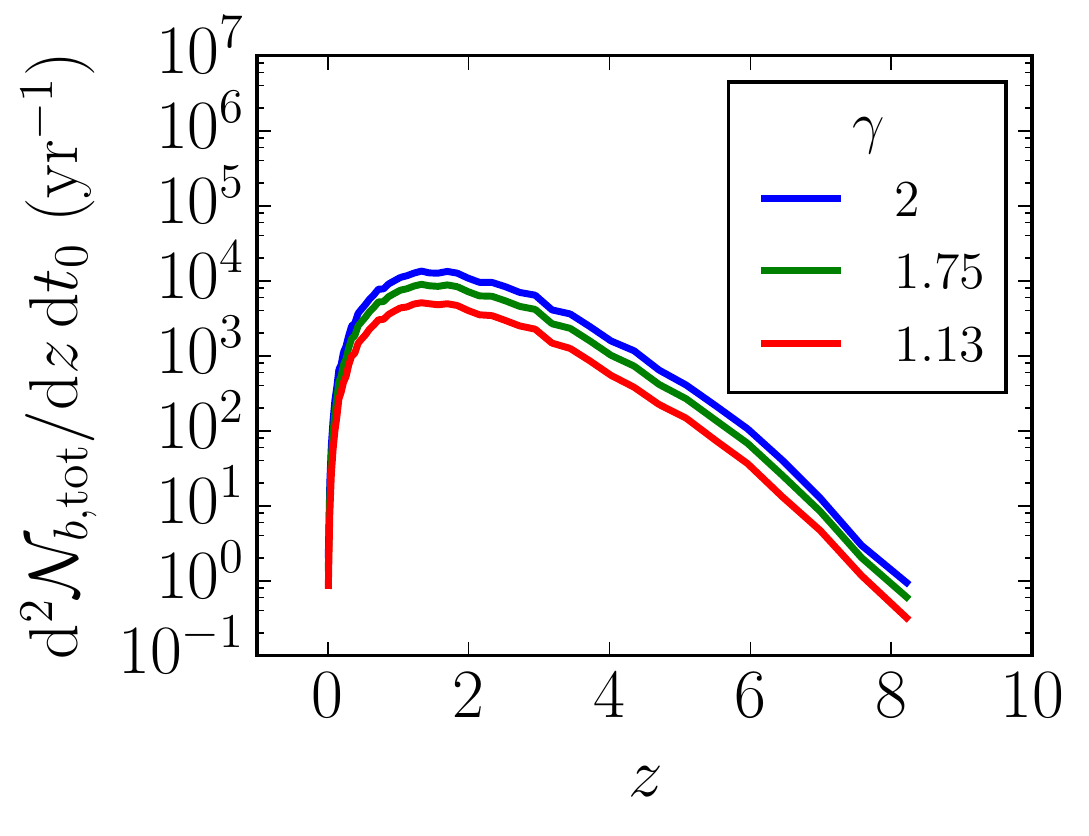}
\caption{Differential MBHB TDE rate as a function of primary black hole mass (left), mass ratio (centre), and redshift (right), in the observer frame. The blue, green and red lines correspond to $\gamma= 2, 1.75, 1.13$ respectively, as labelled in the figure. The shaded area in the central plot marks the region where the TDE rate scalings derived in \protect\cite{chen11}, and used in this work, are more reliably applicable.}
\label{fig:dNdlogTDE}
\end{figure*}

Given the merger rate distribution, we now proceed to calculate the tidal disruption rate. We multiply equation \eqref{eq:d4Ndt0} by $\mathcal{N}_{b}(M_1,q,\gamma)$ as defined in equation \eqref{eq:ntd}, using the values of $M_1$ and $q$ at the bin centres, limiting the mass range to $M_1\in[10^5,6\times10^7]M_\odot$, the upper limit being set to the value of $M_\text{crit}$ given in Section \ref{sec:tdrates}. The lower limit is arbitrarily set around the minimum estimated MBH mass of an observed TDE. This gives the total number of disruptions from MBHBs along the cosmic history:
\begin{equation}
    \frac{\dd[4]{\mathcal{N}_{b,\text{tot}}}}{\dd{\log M_1}\,\dd{\log q}\,\dd{z}\,\dd{t_0}}.
    \label{eq:d4NTD}
\end{equation}
As before, this can be integrated to get one-dimensional distributions, and the global MBHB TDE rate as observed on Earth. In the calculation, we consider $\gamma\in[1,2]$. Results for selected $\gamma$ are shown  in Fig.\ \ref{fig:dNdlogTDE}. MBHB TDEs prefer high mass systems, making the adopted cutoff at $M_1=10^5\msun$ irrelevant. The mass ratio distribution shows a broad peak around $q=0.1$ and the redshift distribution closely follows that of the parent MBHBs. It can be seen from the one-dimensional distributions that there is only a factor of $\sim 3$ difference in the distributions between $\gamma=2$ and $\gamma=1.13$\footnote{The value $\gamma = 1.13$ has been chosen as representative, based on the recent measurement of the slope of the inner stellar cusp in the Milky Way centre \citep{2018A&A...609A..27S}.}, and this is reflected in the global MBHB TDE rate, which we calculate to be $(29989,19834,11152)~\si{yr^{-1}}$ for $\gamma=(2,1.75,1.13)$.
\par
By normalising its integral to unity, we can treat equation \eqref{eq:d4NTD} as a probability distribution, so that we can sample events by selecting a bin based on its relative TDE rate, and by taking that bin's parameters as the parameters of the MBHB producing the event.

\subsection{Single MBH TDE Rate}
The single black hole population is also built from the \texttt{guo2010a} catalogue. We extract all black holes contained in the catalogue which have $z<8$ and $M_{\rm BH}<10^8M_\odot$, leading to a population of $N_s\approx3.3\times 10^7$ black holes.
\par
For each of the simulation's redshift snapshots, we bin the black holes by mass in 30 logarithmically spaced bins for $M_{\rm BH}\in[10^5,6\times10^7]~\msun$. In each bin, we divide the number of MBHs by the simulation volume in order to obtain the differential number density 
\begin{equation}
	\frac{\dd[2]{N_s}}{\dd{\log M_{\rm BH}}\,\dd{V_c}},
\end{equation}
at each redshift. Finally, we compute the distribution of MBHs in mass and redshift as
\begin{equation}
	\frac{\dd[2]{N_s}}{\dd{\log M_{\rm BH}}\,\dd{z}} = \frac{\dd[2]{N_s}}{\dd{\log M_{\rm BH}}\,\dd{V_c}}\cdot\dv{V_c}{z}.
\end{equation}
\par
In order to convert this into a distribution of tidal disruption rates, we use
\begin{equation}
	\frac{\dd[3]{\mathcal{N}_{s,\text{tot}}}}{\dd{\log M_{\rm BH}}\,\dd{z}\,\dd{t_0}} = \frac{\dd[2]{N_s}}{\dd{\log M_{\rm BH}}\,\dd{z}}\cdot\frac{\dot{\mathcal{N}}_{s}}{1+z},
\end{equation}
where $\dot{\mathcal{N}}_{s}(M_{\rm BH})$ is the tidal disruption rate for a single MBH of mass $M_{\rm BH}$, as given in equation \eqref{eq:ntdsinglee}, and the factor of $1+z$ corrects the rate to the $z=0$ frame. As with the binary case, we can integrate this distribution over $M_{\rm BH}$ and $z$ to obtain the global rate of tidal disruptions arising from single galactic centre black holes. This yields a total rate of $\sim8.5\times 10^5$~yr$^{-1}$ for a fiducial mass growth of $0.45M_\odot$ per TDE, and $\sim2\times 10^6$~yr$^{-1}$ for a reduced mass growth of $0.15M_\odot$ per TDE. By comparing these numbers with those obtained in the previous section, it can already be seen that MBHBs can potentially contribute 0.5--4 per cent of the global TDE rate in the Universe.
\par
\begin{figure}
	\centering
	\includegraphics[width=\linewidth]{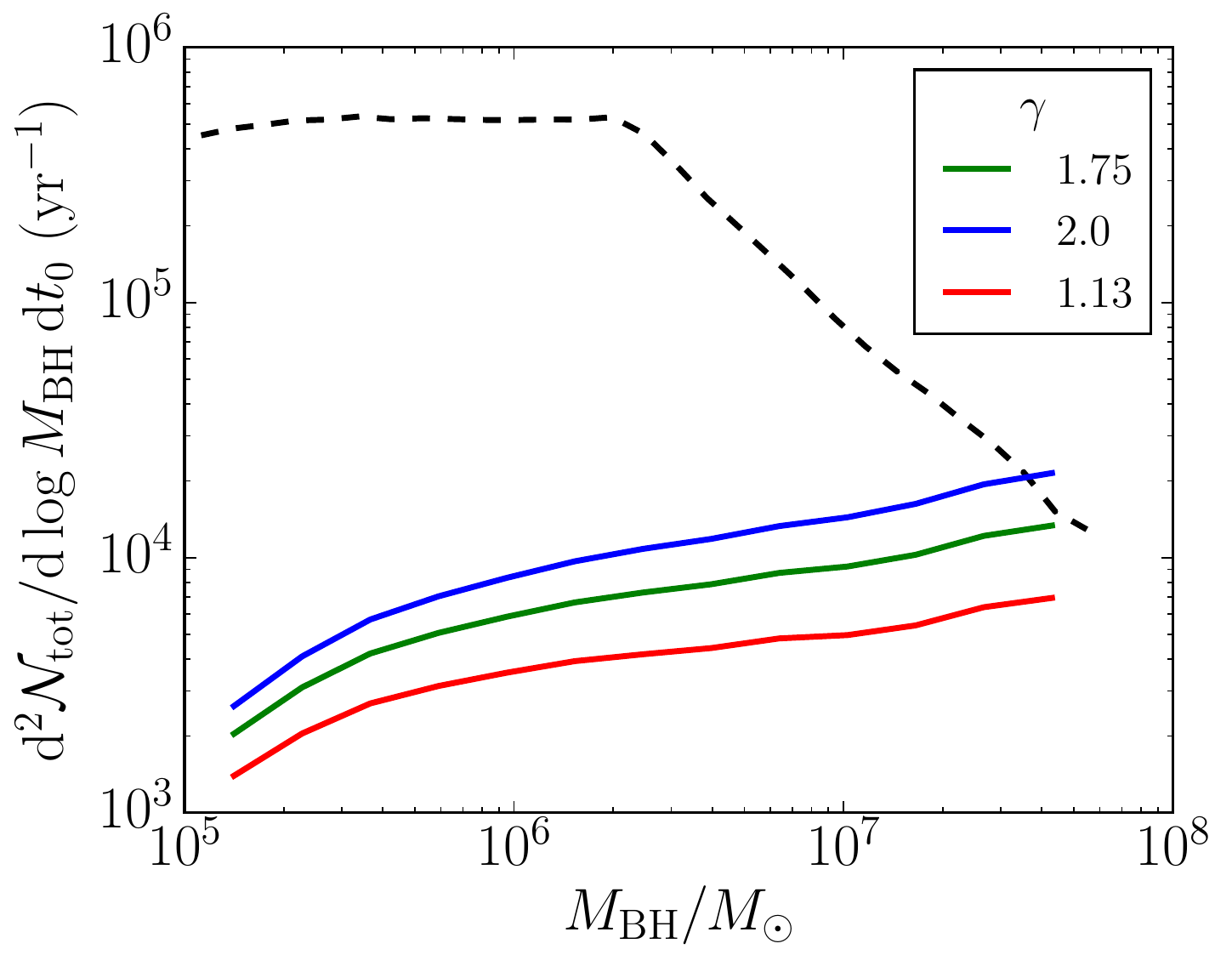}
	\caption{TDE mass functions for MBH (black dashed line) and MBHB (coloured lines) induced disruptions.}
	\label{fig:massfuncs}
\end{figure}

As a point of interest, Fig.\ \ref{fig:massfuncs} compares the differential mass functions of TDE rates due to MBHs and MBHBs. Contrary to the MBH case, the number of MBHB TDEs is an increasing function of the MBH mass. This behaviour arises mainly from the mass scaling of equation \eqref{eq:ntd}. Also noticeable is the flattening of the single MBH TDE mass function for $M_{\text{BH}}<10^6\msun$, which is due to the imposition of a maximum allowed MBH mass growth as per equation \eqref{eq:ntdsinglee}.

\section{TDE Emission}
\label{sec:emission}

\begin{figure*}
\begin{tabular}{cc}
\includegraphics[width=0.5\linewidth]{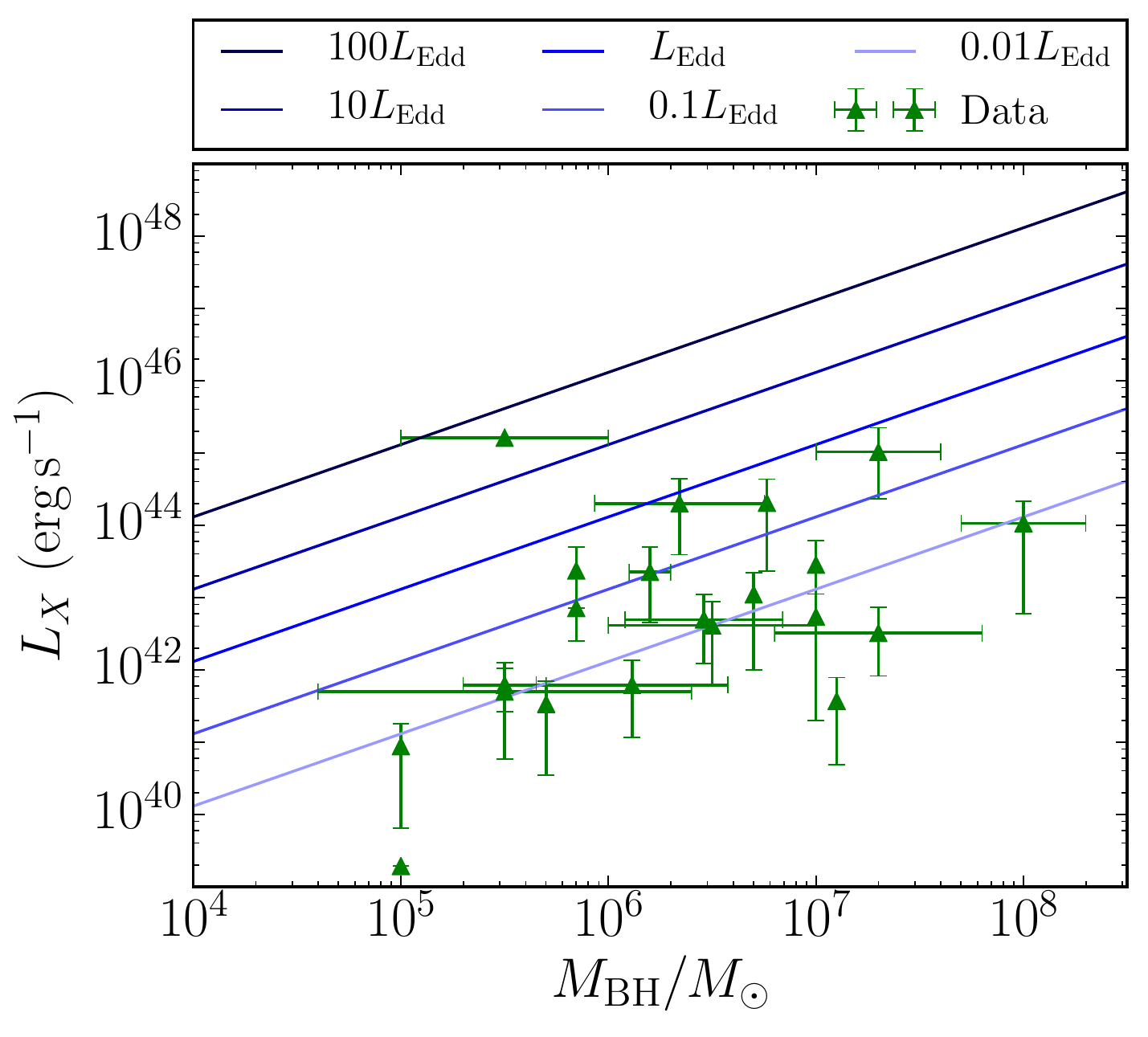}&
\includegraphics[width=0.5\linewidth]{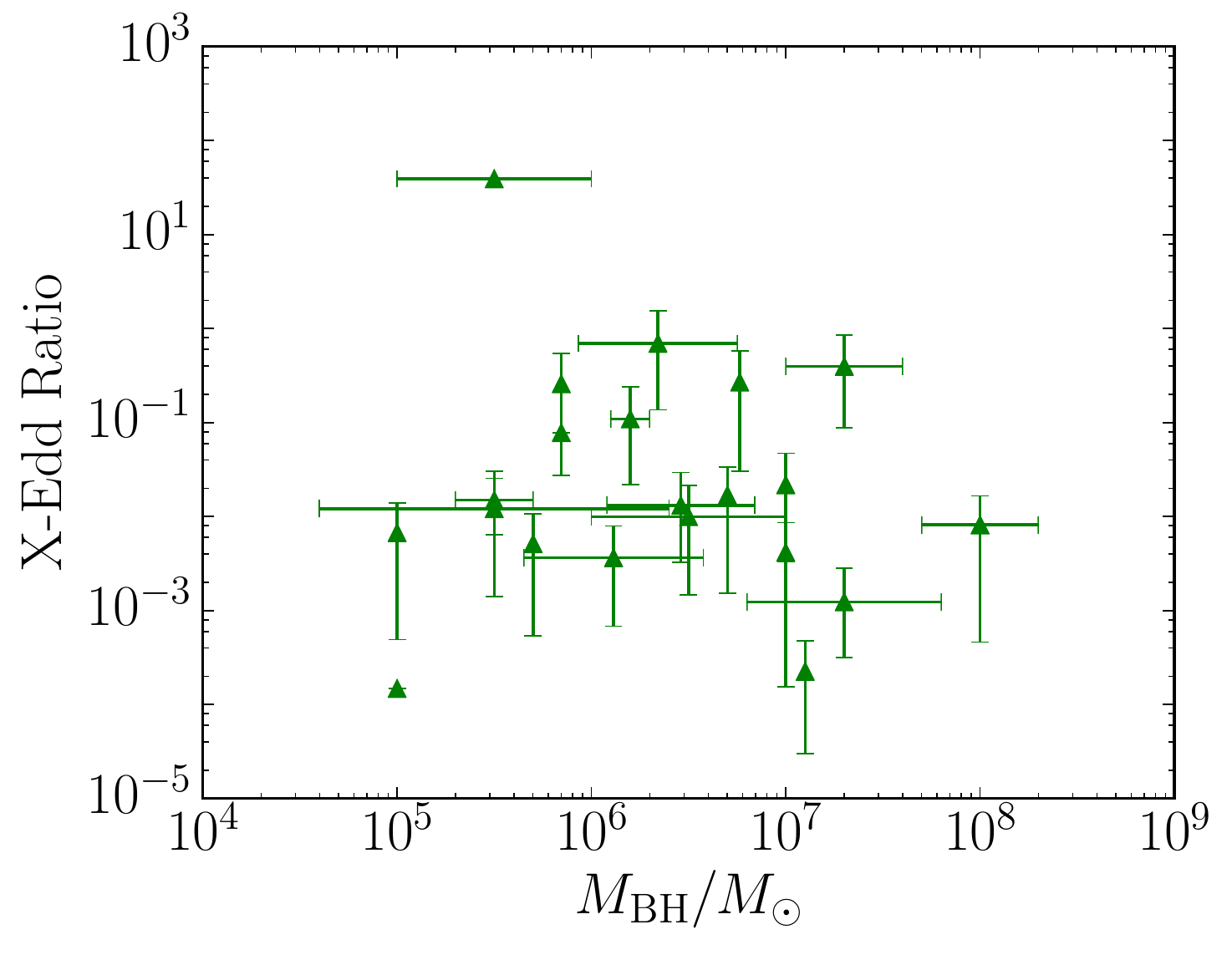}
\end{tabular}
\caption{Left panel: X-ray luminosity against black hole mass for data from \citet{auchettl17}, compared with factor-of-10 multiples of the Eddington luminosity (solid lines). Right panel: derived X--Edd ratio against black hole mass for the same set of observations.}
\label{fig:LX}
\end{figure*}

TDEs candidates have been recorded at different wavelengths, including optical, UV and X-ray. In all bands, these events are quite bright, with inferred isotropic luminosities in the range $10^{42}$--$10^{44}$~erg s$^{-1}$. Providing a coherent physical explanation to such diverse phenomenology has proven difficult, and a comprehensive theoretical model is still missing. As standard accretion disc emission can hardly reproduce observations, optical spectra are tentatively explained with radiation-driven winds \citep{strubbe09,lodatorossi11}, inflation of infalling gas into an envelope \citep{coughlin14}, or shocks of the self-intersecting disrupted debris prior to circularisation \citep{piran15}. The scarcity of TDEs observed at multiple wavelengths suggests that the energy of the observed photons might be related either to the time at which the event is caught or to orientation effects. The first option has been suggested by observations of the TDE candidate ASASSN-15oi, for which \citet{gezari17} report an X-ray brightening coincident with fading UV emission \citep[see also][]{2018MNRAS.480.5689H}. The latter has been recently proposed by \cite{2018ApJ...859L..20D}. In their 3D GRMHD simulation coupled with Monte--Carlo radiative transfer, they find that the resulting angle-dependent outflow produces radiation peaking at different wavelengths depending on the inclination angle to the observer.

The construction of a self-consistent physical model for TDE spectra is beyond the scope of our paper. With the aim of predicting observation rates for future surveys, we therefore create an ad-hoc phenomenological spectrum with physically motivated components that can reproduce observational data. Despite our focus on MBHB systems, we model all accretion as being only around a single MBH of mass $M_{\rm BH}$.

\subsection{Eddington Accretion Disk}

\begin{figure}
    \centering
    \includegraphics[width=\linewidth]{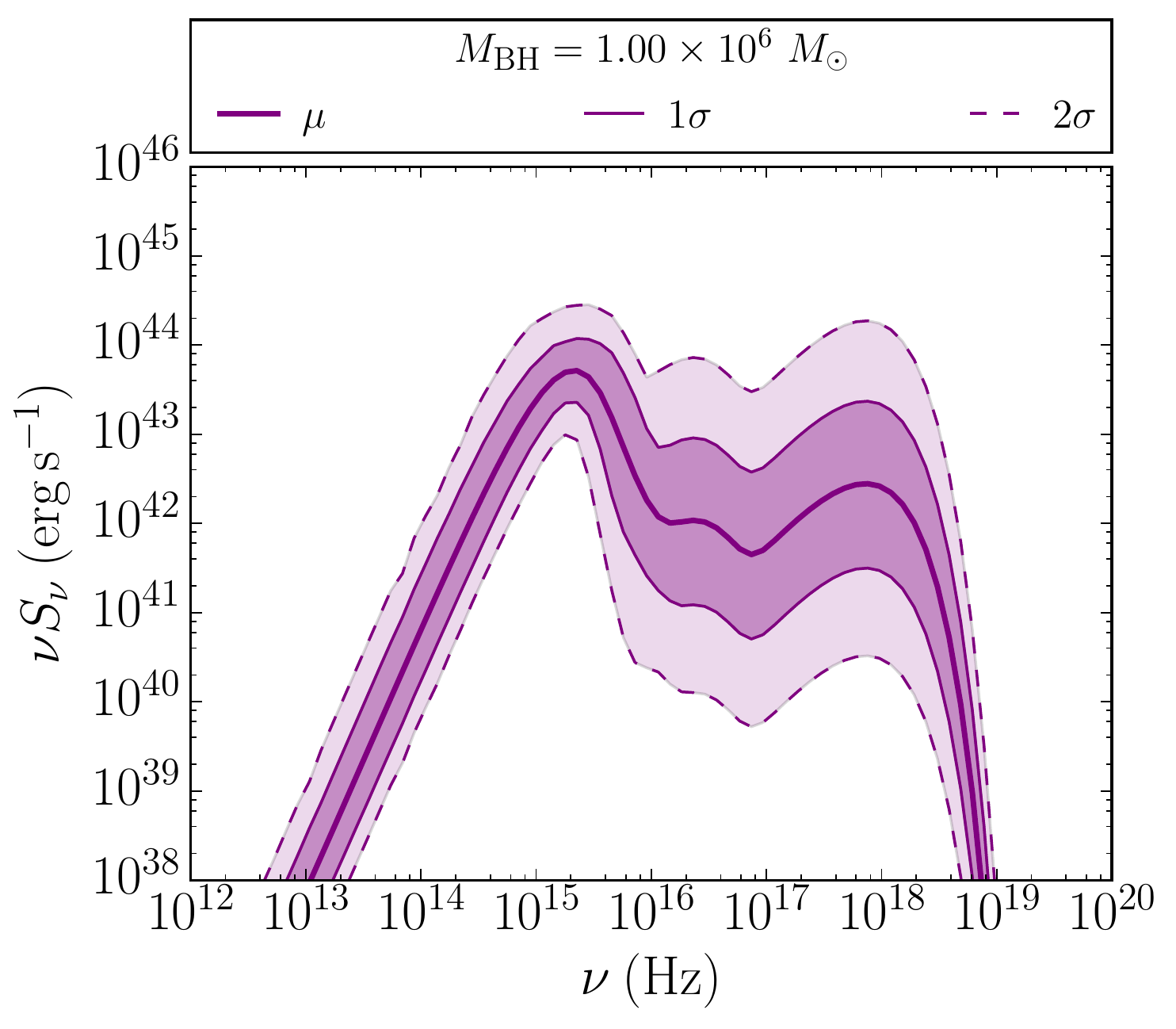}
    \caption{Phenomenological spectrum for a TDE produced by a $10^6M_\odot$ black hole. The thicker central curve shows the `mean' spectrum, $\mu$, for such an event, with $1\sigma$ and $2\sigma$ confidence intervals showing the typical range of spectra our model would produce.}
    \label{fig:SnuPhenom}
\end{figure}
\begin{figure*}
    \centering
    \includegraphics[width=\textwidth]{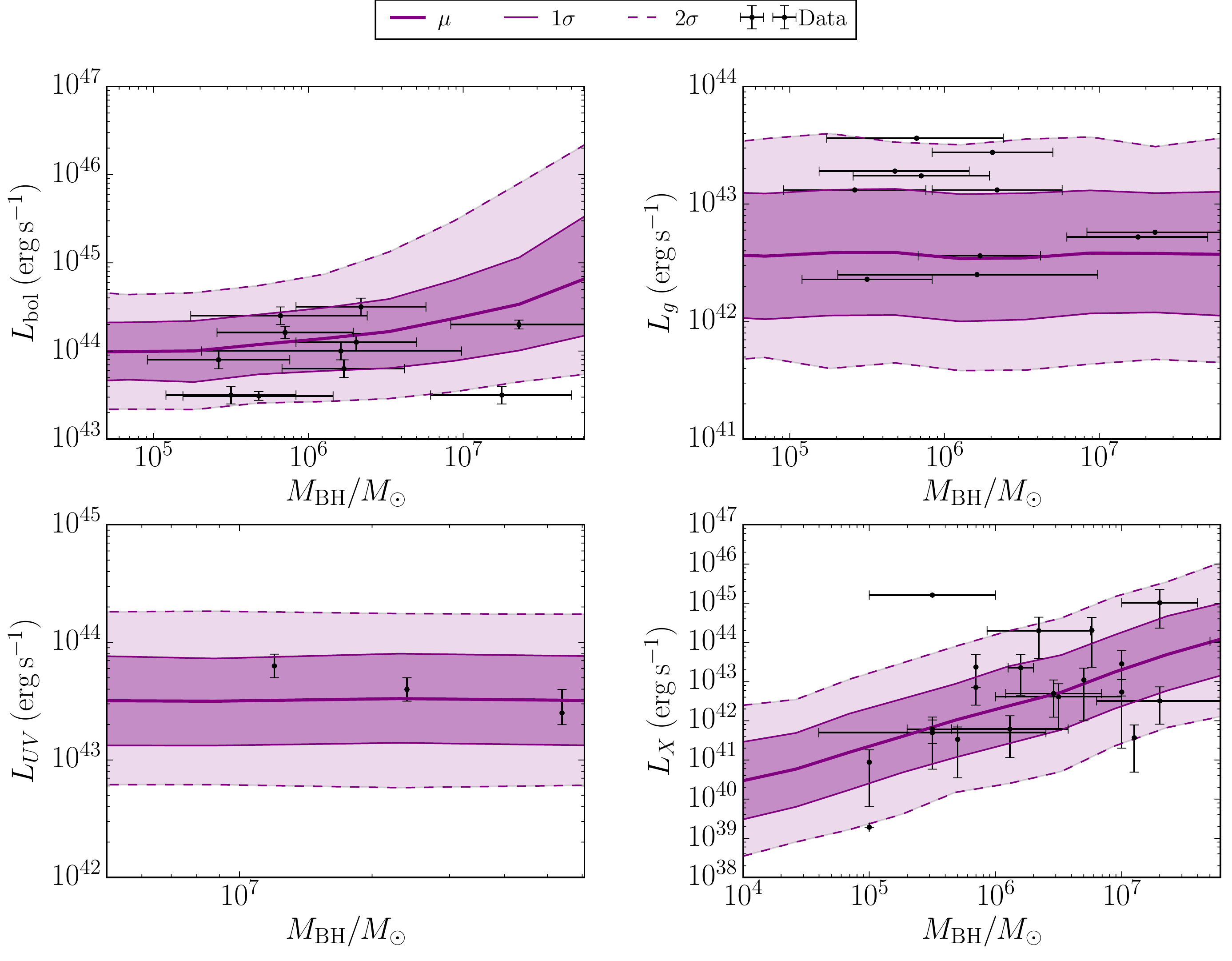}
    \caption{Luminosities predicted by our phenomenological model, compared to the observed luminosities of past disruptions. The bolometric and $g$-band data are taken from \protect\citet{wevers17}, the X-ray data are collated from \protect\citet{auchettl17} and the UV data are from three \textit{GALEX} observations \protect\citep{gezari06,gezari08,gezari09}. We calculate $L_g$ in \textit{LSST}'s $g$-band (3877--5665~\AA, see Table \ref{tab:filternumbers}), $L_{UV}$ in the combined \textit{NUV+FUV} range of \textit{GALEX} \protect\citep[1344--2831~\AA, following][]{gezari06}, and $L_X$ in the range 0.3--2~keV \protect\citep[as used by][]{auchettl17}.}
    \label{fig:bandLums}
\end{figure*}

We take the limits of the accretion disc as those defined by \citet{lodatorossi11}:
\begin{align}
    r_\xin&=3r_S=\frac{6G\mbh}{c^2}, \\
    r_\xout&=\frac{2r_t}{\beta}.
\end{align}
Here $r_S$ is the Schwarzschild radius of the black hole, and $\beta$ is a `penetration factor' defined as the ratio of the pericentre of the star to the tidal radius. For simplicity, we take $\beta=1$ and assume solar-type stars for the calculation of $r_t$ (using equation \eqref{eq:rt}). The previously mentioned critical mass $M_\text{crit}$ is the mass at which $r_\xin=r_\xout$, given these assumptions\footnote{We note that for highly spinning MBHs the prograde innermost stable circular orbit is located much closer to the horizon; as a result, these black holes of up to a ${\rm few }\times 10^8M_\odot$  can support TDEs \citep{2012PhRvD..85b4037K}. Although this might enhance the rate of tidal disruptions, we do not consider spin in our calculation, making our results conservative in this respect.}.

The base of our phenomenological model follows the classic description of an accretion disc by \cite{pringle81}, in which each annulus of the disc radiates as a black body with a different temperature. The temperature of the disc is given by
\begin{equation}
    T(r)=\left[\frac{3c^2\,r_{S,\odot}\mdbh}{16\pi r^3\sigma}\pfrac{\mbh}{\msun}\left(1-\sqrt{\frac{r_\xin}{r}}\right)\right]^{1/4},
    \label{eq:Ts}
\end{equation}
where $r_{S,\odot}$ is the Schwarzschild radius of the sun, $\mdbh$ is the accretion rate of the black hole, and $\sigma$ is the Stefan--Boltzmann constant. In our model, we assume the peak accretion rate of the black hole to be equal to the Eddington accretion rate, given by \cite{lodatorossi11} as
\begin{equation}
    \mdedd=1.3\e{18}\,\frac{\mbh}{\msun}\pfracp{\eta}{0.1}{-1}\,\si{g.s^{-1}},
    \label{eq:mdotedd}
\end{equation}
where $\eta$ is the radiative efficiency of the accretion process, which we take as 1/12.

At each radius, the spectrum emitted is a black-body
\begin{equation}
    B[\nu,T(r)]=\frac{2h}{c^2}\frac{\nu^3}{e^{h\nu/k_B T(r)}-1},
\end{equation}
where $\nu$ is frequency, $h$ is Planck's constant, and $k_B$ is the Boltzmann constant. Integrating over all radii gives the total spectrum for the disc,
\begin{equation}
    S(\nu)=2\pi\int_{r_\xin}^{r_\xout} B[\nu,T(r)]\,2\pi r\dd{r}.
\end{equation}
Here, the factor of 2 accounts for both sides of the disc, and $\pi$ accounts for the integration of emission over all solid angles.

When integrated over all frequencies, this spectrum yields the Eddington luminosity,
\begin{equation}
    L_\edd=1.3\e{38}\pfracp{\eta}{0.1}{-1}\pfrac{M_{\rm BH}}{\msun}.
\end{equation}

When including the Pringle--Eddington spectra in our phenomenological model, we multiply them by an additional factor, the `X--Edd ratio,' so that the bolometric luminosity will not always equal the Eddington luminosity exactly. This X--Edd ratio is described in Section \ref{sec:phenomX}.

\subsection{Optical Black Body}

The Pringle--Eddington emission model above peaks in the soft X-ray range, and drops off significantly for low frequencies. However, TDEs have been observed in the optical band, suggesting another mechanism contributing to the emission. 

\cite{wevers17} give black-body fits for a number of optical observations; the corresponding temperatures are typically an order of magnitude lower than the range of temperatures given by equation \eqref{eq:Ts}. We therefore add an extra component to the base spectrum, corresponding to a lower temperature black-body. Temperatures are selected independently of mass, from a uniform distribution between $1.2\times10^4$~K and $4.9\times10^4$~K -- these are the lowest and highest temperatures given by \cite{wevers17}. The black-body spectrum for a selected temperature is normalized such that it integrates to a particular luminosity, which is sampled from a Gaussian derived from the bolometric luminosities given by \cite{wevers17}.

\subsection{X-Ray Synchrotron Emission}
\label{sec:phenomX}

The Pringle--Eddington model spectrum typically peaks in the soft X-ray region, but drops off rapidly beyond the peak. This implies a low integrated luminosity in the X-ray, which does not fit observational data. We therefore need to add another component to our model to increase the X-ray luminosity.

Many observational papers fit a number of different models to their observed spectra. In some cases, a black-body fit is best; however, these fits typically have black-body temperatures within the Pringle temperature range, so adding a black-body bump to the Pringle--Eddington spectrum would not have a significant effect. In other cases, a power-law fit seems more suitable, corresponding to inverse Compton scattering from a corona around the black hole. \cite{auchettl17} collate data on a number of observations, and fit their spectra in the 0.3--2~keV band, providing power-law fits and X-ray luminosities for TDE candidates. We tested a power-law addition to the Pringle--Eddington spectrum, with slope indices sampled based on the fits in \cite{auchettl17} (we use their `X-Ray TDE, `Likely X-Ray TDE' and `Possible X-Ray TDE' categories, excluding any jetted events). However, this was insufficient to recreate the luminosities given in their paper. This was mainly due to our imposed constraint that the power law addition must start after the peak of the Pringle--Eddington spectrum and join with it in a continuous manner. 

\citet{auchettl17} also try a fit of the form $\nu S(\nu)\propto\nu^{1/2}e^{\nu/\nu_\text{cut}}$, where $\nu_\text{cut}$ is a `cutoff' frequency. This corresponds to X-ray synchrotron emission. We take $\nu_\text{cut}$ to be 2~keV, the upper limit of the band used by \cite{auchettl17}, and find that this produces a peak in the spectrum at higher frequency than the Pringle--Eddington model, making it a suitable addition to our phenomenological model.

To normalize this component, we use luminosities given by \cite{auchettl17}, and black hole masses collated from the original observational papers for the same events \citep[as referenced by][]{auchettl17}. We notice that the X-ray luminosity seems to increase linearly with mass (see Fig.\ \ref{fig:LX}, left panel), and therefore define an `X--Edd ratio' as the ratio of the X-ray luminosity to the Eddington luminosity (Fig.\ \ref{fig:LX}, right panel). The data fall in a roughly Gaussian manner around the $0.01L_\edd$ line, with the exception of the point that lies between the $10L_\edd$ and $100L_\edd$ lines in the left panel of Fig.\ \ref{fig:LX}. The point belongs to the event 3XMM J152130.7+074916, of which only one observation has been made. \cite{auchettl17} derive an X-ray luminosity two orders of magnitude higher for this event than is given in the original observational paper \citep{lin15}. Due to this discrepancy, we leave this event out of our calculations, and create a Gaussian distribution using the mean and standard deviation of the remaining data. The synchrotron emission `bump' is normalized to a luminosity sampled from this distribution.

Fig.\ \ref{fig:SnuPhenom} shows the range of typical spectra which our phenomenological model predicts for TDEs arising from a $10^6M_\odot$ MBH. Fig.\ \ref{fig:bandLums} compares the luminosities predicted by the model across the relevant parts of the EM spectrum to the observed luminosities of past events. The fit to the observed data is satisfactory in all bands, which is expected, given the phenomenological nature of the model. Although the emission model does not try to connect the different spectral components to the underlying physics of the disruption, a solid match of the luminosity in different bands as a function of the system parameters (essentially the MBH mass) should be all that is needed to make reliable predictions for future surveys. Nonetheless, in Appendix \ref{app:emissionmodels} we speculate as to the impact of adopting some of the emission models which exist in the literature.

\subsection{Time Variance}
Despite the time dependent evolution of a TDE likely depending on the band at which the event is observed \citep[see e.g.][]{lodatorossi11}, we take the simplifying assumption that the light curve of a TDE decays as a $t^{-5/3}$ power law
\begin{equation}
    L(t)=L(0)\pfracp{t+\tau}{\tau}{-5/3},
\end{equation}
where $\tau$ is a characteristic time-scale.

In the X-ray band, observational papers give the factor by which the luminosity decays over a particular length of time. If measurements are made at $t=0$ and $t=t'$, and if $L(t')=L(0)/K$, where $K$ is some constant, then the time-scale is given by
\begin{equation}
    \tau=\frac{t'}{K^{3/5}-1}.
\end{equation}
Our X-ray decay factors and corresponding time-scales are collated in Table \ref{tab:XrayTaus}. The values of $\tau$ span two orders of magnitude; this is probably due to the specific parameters of each event, such as the type of star and the value of $\beta$, which we do not model here.
\begin{table}
    \centering
	\caption{X-ray event decay time-scales. $K$ is the factor by which the luminosity has decreased at time $t'$.}
	\begin{threeparttable}
		\label{tab:XrayTaus}
		\begin{tabular}{c c c C}\toprule
		     Event & $K$ & $t'$ & \tau~(\text{s})\\ \midrule
		     NGC 5905\tnote{a,b} & 80 & 2\,yr & 4.9\e{6}\\
		     RXJ1242.6-1119\tnote{c} & 20 & 1.5\,yr & 9.4\e{6}\\
		     RXJ1420.4+5334\tnote{d} & 150 & 0.5\,yr & 8.2\e{5} \\
		     NGC 3599\tnote{e} & 30 & 3\,yr & 1.4\e{7} \\
		     SDSS J132341.97+482701.3\tnote{e} & 40 & 3\,yr & 1.2\e{7} \\
		     SDSS J120136.02+300305.5\tnote{f} & 300 & 300\,d & 8.7\e{5} \\
		     TDXF 1347-3254\tnote{g} & 650 & 13\,yr & 8.5\e{6} \\
		     SDSS J131122.5-012345.6\tnote{h} & 30 & 2\,yr &9.4\e{6} \\
		     WINGS J1348\tnote{i} & $10^4$ & 15\,yr & 1.9\e{6} \\
		     RBS 1032\tnote{j} & 100--300 & 14\,yr &\sim 1.9\e{7} \\ \bottomrule
		\end{tabular}
		\begin{tablenotes}
			\begin{multicols}{2}
				\item [a] \citet{bade96}
				\item [b] \citet{komossabade99}
				\item [c] \citet{komossagreiner99}
				\item [d] \citet{greinerschwarz00}
				\item [e] \citet{esquej08}
				\item [f] \citet{saxton12}
				\item [g] \citet{cappelluti09}
				\item [h] \citet{maksym10}
				\item [i] \citet{maksym14a}
				\item [j] \citet{maksym14b}
			\end{multicols}
		\end{tablenotes}
	\end{threeparttable}
\end{table}

For optical events, \cite{wevers17} provide the decay rate in magnitudes per 100 days. If this rate is $\kappa$ magnitudes per 100 days, then we find that
\begin{equation}
    \tau=\frac{t'}{10^{24\kappa/25}-1},
\end{equation}
for comparison to X-ray time-scales.

\begin{figure}
    \centering
    \includegraphics[width=\linewidth]{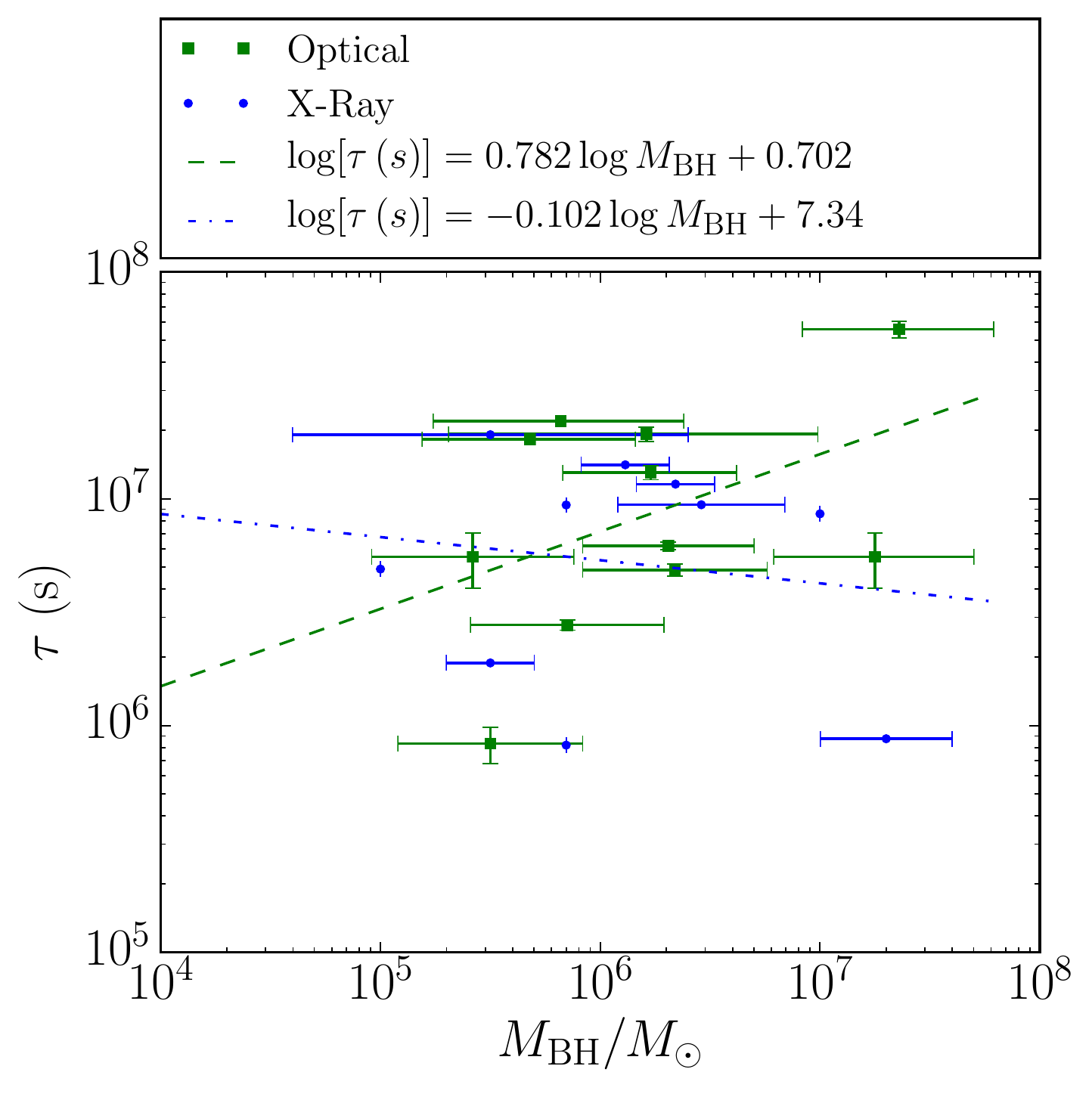}
    \caption{Decay time-scales against mass, for optical (green) and X-ray (blue) events.}
    \label{fig:t0VsMBH}
\end{figure}

As mentioned previously, \cite{gezari17} suggest that the optical and X-ray components of TDE flares are produced by different mechanisms, peaking at different times and possibly decaying with different time-scales. Our Fig.\ \ref{fig:t0VsMBH} tentatively supports this suggestion; the relation between time-scale and black hole mass appears to follow different trends in the different regions. We do not investigate this further in this paper, however.

\section{Detectors}
\label{sec:detectors}
We now turn our attention to selected current and future survey instruments suited to TDE identification. \textit{LSST} \citep{2009arXiv0912.0201L} and \textit{eROSITA} \citep{erosita12} will become operational in the next few years, with the former covering the optical sky from the ground, and the latter covering the X-ray sky from space. These are our main focus, but we also consider \textit{Gaia}, a currently operational satellite of the European Space Agency which, although not primarily aimed at detecting transients, has the potential to make several TDE detections, due to its reasonable cadence and whole-sky coverage \citep[e.g.][]{blagorodnova16}. In this section, we discuss our methods for calculating the rates of tidal disruptions which these surveys will be able to detect, both from MBHBs and single MBHs. Although these surveys have somewhat different observing strategies, we adopt the same criteria for a good detection in all cases -- the observation of a three point light curve in which at least 2~mag of decay is visible.

\subsection{\textit{LSST}}
\label{sec:detectors-LSST}
\textit{LSST} is a ground based optical survey due to begin observing in 2022, from which time it will spend 10 years regularly scanning the sky using six photometric filters. Due to the large-scale, repeated sky coverage, \textit{LSST} is ideal for transient detection. We consider here the \texttt{minion\_1016} observing strategy\footnote{Although \texttt{minion\_1016} was the baseline strategy when \protect\citet[Version 1.0]{LSSTwhitepaper} was released, this is still subject to change, with the operations simulation having been subsequently revised, and proposals for alternative cadence strategies currently being taken \protect\citep[see][]{baseline2018a,cadencecall}. Changes in filter sensitivities between \texttt{baseline2018a}/\texttt{astro-lsst-01\_2022} \protect\citep{baseline2018a} and \texttt{minion\_1016} are insignificant, and the cadence sensitivity of our model is not high, so \texttt{minion\_1016} is sufficient for our purposes.}, which gives an average cadence of between 20 days ($r$-band) and 30 days ($u$-band) \citep[][\S\,2.3, see also \S\,6.6.1--6.6.2 for the TDE science case]{LSSTwhitepaper}. Adopting this as our model for \textit{LSST}'s operations, we assume a worst case cadence of $t_{\text{cadence}}=30$~days in all filters, and adopt the single-visit magnitude thresholds for each filter which are given in the \texttt{minion\_1016} strategy. These are listed, along with the frequency ranges of each filter in Table \ref{tab:filternumbers}.
\begin{table*}
	\centering
	\caption{Wavelength and frequency limits for \textit{LSST}'s optical filters, along with their single visit magnitude thresholds, $m_\phi$. Wavelength limits are taken from the SVO Filter Profile Service \citep[\protect\url{http://svo2.cab.inta-csic.es/theory/fps/},][]{rodrigo12,rodrigo13}. Magnitude thresholds are taken from the former \textit{LSST} baseline strategy, \texttt{minion\_1016} \citep[][\S\,2.3]{LSSTwhitepaper}.}
	\label{tab:filternumbers}
	\begin{tabular}{c c c c c c} \toprule
		Filter, $\phi$ & $\lambda_{\phi,\text{min}}$~(\AA) & $\lambda_{\phi,\text{max}}$~(\AA) & $\nu_{\phi,\text{min}}$~($10^{14}$~Hz) & $\nu_{\phi,\text{max}}$~($10^{14}$~Hz) & $m_\phi$~(mag)\\ \midrule
		$u$ & 3182 & 4082 & 7.349 & 9.428 & 23.9\\
		$g$ & 3877 & 5665 & 5.296 & 7.738 & 25.0\\
		$r$ & 5375 & 7055 & 4.252 & 5.581 & 24.7\\
		$i$ & 6765 & 8325 & 3.664 & 4.435 & 24.0\\
		$z$ & 8035 & 9375 & 3.200 & 3.734 & 23.3\\
		$y$ & 9089 & 10859 & 2.763 & 3.301 & 22.1\\
		\bottomrule
	\end{tabular}
\end{table*}
\par
\begin{figure}
	\centering
	\includegraphics[width=\linewidth]{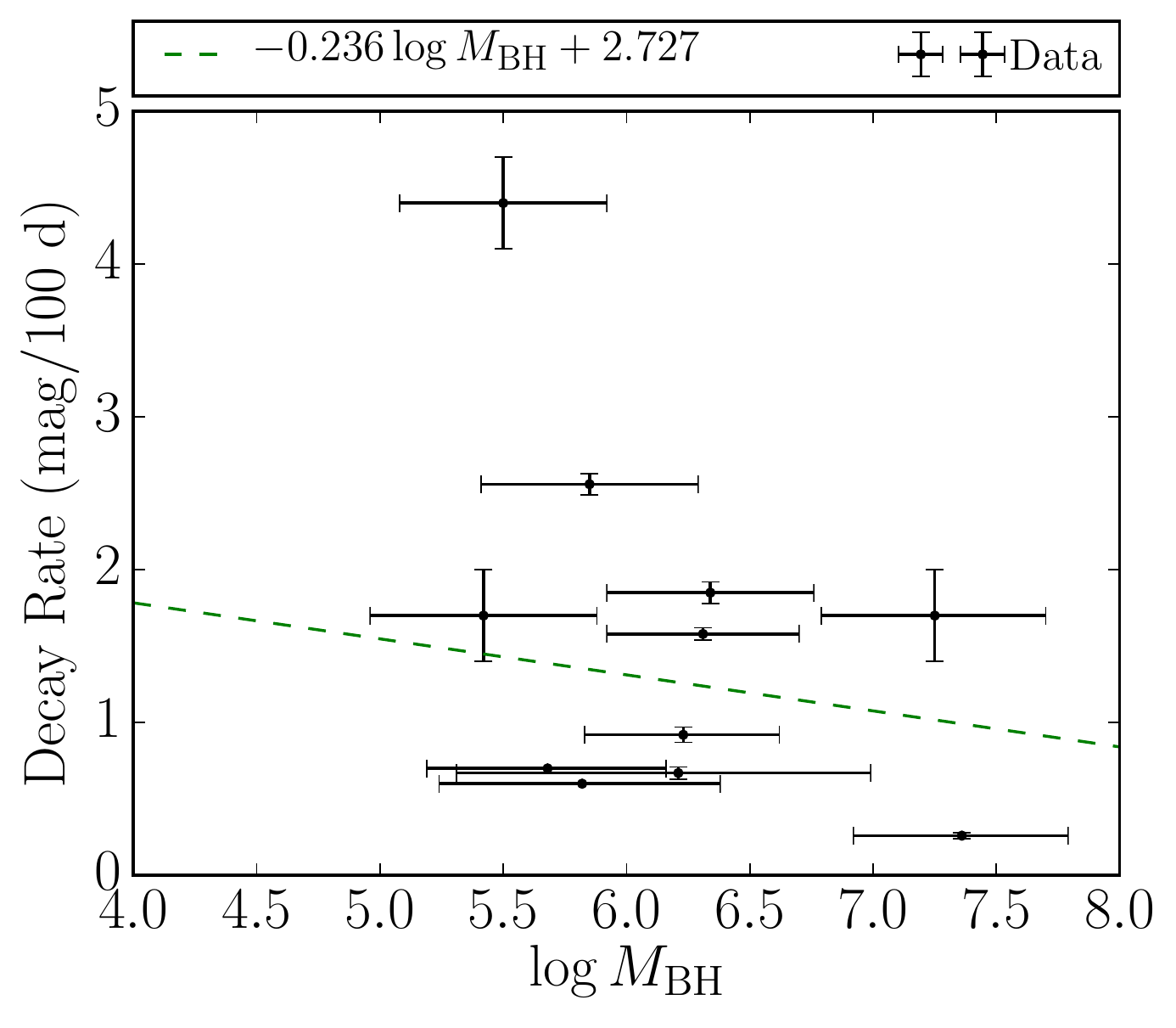}
	\caption{Event decay rates taken from \citet{wevers17}, along with the regression line from which equation (\ref{eq:m100d}) is derived. The regression line is not fit to the outlying point with a decay rate of 4.4~$\text{mag}/100$~d.}
	\label{fig:decayrates}
\end{figure}

We additionally impose the requirement that a good \textit{LSST} detection is one which is observable in at least three filters. For a particular filter, $\phi$, we find the maximum redshift at which a TDE could be observed, $z_\text{max}(M_\text{\rm BH})$, by solving the equation
\begin{multline}
	(1+z_\text{max})\left(\frac{D_H}{\text{pc}}\right)\int^{z_\text{max}}_0\frac{\mathrm{d}z'}{E(z')} \\= 10^{[m_\phi-M_\phi(M_{\rm BH},z_{\text{max}})+5]/5},
\end{multline}
for $z_{\text{max}}$, across a range of black hole masses. The left hand side here comes from the standard expression for luminosity distance as a function of redshift, whilst the right hand side comes from relating distance modulus to luminosity distance \citep[e.g.][]{hogg99}. On the right hand side, $m_\phi$ is set to the magnitude threshold of the particular filter, being considered, and $M_\phi(M_{\rm BH},z_{\text{max}})$ is the peak magnitude an event would require in that filter to meet our criteria for a good detection. This is given by
\begin{multline}
	M_\phi = -\frac{5}{2}\log\left(\frac{L_\phi(M_{\rm BH},z_\text{max})}{3.02\times10^{35}~\text{erg\,s}^{-1}}\right) \\ + \max[2,M_\text{decay}(M_{\rm BH})].
	\label{eq:absolutemagwithmax}
\end{multline}
Here, the first term is the standard luminosity--magnitude conversion,
where $L_\phi$ is the redshift-corrected luminosity of the event in the filter $\phi$, given by
\begin{equation}
	L_\phi = \int^{(1+z_\text{max})\nu_{\phi,\text{max}}}_{(1+z_\text{max})\nu_{\phi,\text{min}}}S_\nu(\nu)\,\mathrm{d}\nu.
\end{equation}
In this expression, $S_\nu$ is defined using our phenomenological model, which is described in Section \ref{sec:emission}, with $\nu_{\phi,\text{max}}$ and $\nu_{\phi,\text{min}}$ taken from Table \ref{tab:filternumbers}. The second term in equation \eqref{eq:absolutemagwithmax} encodes our criteria for a good detection. It requires that the event peaks with sufficient magnitude that the filter can capture 2~mag of decay, and that three data points can be captured before the event decays to below the filter's threshold. This second requirement is encoded in the $M_\text{decay}(M_{\rm BH})$ term, which is given by
\begin{equation}
	M_\text{decay} = \frac{2t_\text{cadence}+t_1}{100~\text{d}}M_{100\text{d}}(M_{\rm BH}).
\end{equation}
Here, $t_1$ is the number of days after peak that the first observation is taken, sampled from a uniform distribution of the form $U(0,t_\text{cadence})$. The quantity $M_{100\text{d}}(M_{\rm BH})$ defines the number of mag the event will decay by over 100~days, sampled from a normal distribution of the form
\begin{equation}
	M_{100\text{d}}(M_{\rm BH})=N[-0.236\times\log M_{\rm BH}+2.73,\,0.71].
	\label{eq:m100d}
\end{equation}
This is derived by fitting a regression line to the decay rates given by \citet{wevers17} (see Fig.\ \ref{fig:decayrates}) and by assuming a Gaussian scatter about this line. The standard deviation is equal to the standard deviation of the residuals.
\par
Due to the random nature of $M_{100\text{d}}(M_{\rm BH})$, $S_\nu$ and $t_1$, there is no one-to-one correspondence between $M_\text{\rm BH}$ and $z_\text{max}(M_\text{\rm BH})$. Rather, for any given filter, there is a probability distribution that the maximum redshift at which an event could be observed lies at $z$. We assume this distribution to be described by a Gaussian function

\begin{equation}
	f_\phi(z) = \frac{1}{\sigma_\phi\sqrt{2\pi}}\exp\left[-\frac{(\log z-B_\phi)^2}{2\sigma_\phi^2}\right].
	\label{eq:fmz}
\end{equation}
The fitting parameters, $B_\phi$ and $\sigma_\phi$, are derived for each filter by generating 100 $z_\text{max}(M_{\rm BH})$ curves for 100 values of $M_{\rm BH}$, and then using a maximum likelihood estimator approach to find the best values. The values derived in this way are included in Table \ref{tab:fittingparams}. Although $f_\phi(z)$ should in principle be a function of $M_{\rm BH}$, we find that in the optical regime, the mass dependence of $z_\text{max}$ is negligible, making $f_\phi(z)$ similarly mass independent. 
\begin{table}
	\centering
	\caption{Fitting parameters for \textit{LSST} for $f_\phi(z)$ (equation \ref{eq:fmz}).}
	\label{tab:fittingparams}
	\begin{tabular}{c c c}\toprule
		Filter, $\phi$ & $B_\phi$ & $\sigma_\phi$ \\ \midrule
		\textit{LSST}-$u$ & -0.80938 & 0.22947\\
		\textit{LSST}-$g$ & -0.54961 & 0.24996\\
		\textit{LSST}-$r$ & -0.82860 & 0.27028\\
		\textit{LSST}-$i$ & -1.17611 & 0.28004\\
		\textit{LSST}-$z$ & -1.55169 & 0.28277\\
		\textit{LSST}-$y$ & -1.73100 & 0.28748\\
		\bottomrule
	\end{tabular}
\end{table}
\par
It follows from equation \eqref{eq:fmz} that the probability of an event at redshift $z$ being beyond the threshold of $\phi$ is simply given by the cumulative distribution of $f_\phi(z)$. Taking the complement of this will thus give the probability that an event at redshift $z$ will be observable in $\phi$. This completeness function has the form
\begin{equation}
	\mathcal{F}_\phi(z) = \frac{1}{4}\left[1-\erf\left(\frac{\log z-B_\phi}{\sigma_\phi\sqrt{2}}\right)\right],
	\label{eq:Fmz}
\end{equation}
where an extra factor of $1/2$ is introduced to account for the fact that $\textit{LSST}$ only covers the southern sky. For each of \textit{LSST}'s filters, $\mathcal{F}_\phi(z)$ is plotted in Fig.\ \ref{fig:observabilitycurves}.

\subsection{\textit{Gaia}}
\label{sec:detectors-gaia}
The satellite \textit{Gaia} \citep{perryman01,2016A&A...595A...1G} began observing in 2014, and has since been undertaking a 5 year mission primarily aimed at mapping the positions and motions of stars in our galaxy. To achieve this, however, multiple observations of each target are required, making \textit{Gaia} naturally suited to detecting transient events, including TDEs \citep[\textit{Gaia}'s transient detection capabilities are discussed thoroughly in][]{blagorodnova16}. When considering \textit{Gaia}, we use the same methodology as with \textit{LSST}. We use a worst case cadence of $t_\text{cadence}=63$~days, which is equal to the rotation period in the instrument's slower scanning direction, and follow \citet{blagorodnova16} in using a magnitude threshold in the $G$ filter of $M_{G}=19$. This filter is a wide optical filter sensitive over 3300--10500\AA.
\par
\begin{figure}
	\centering
	\includegraphics[width=\linewidth]{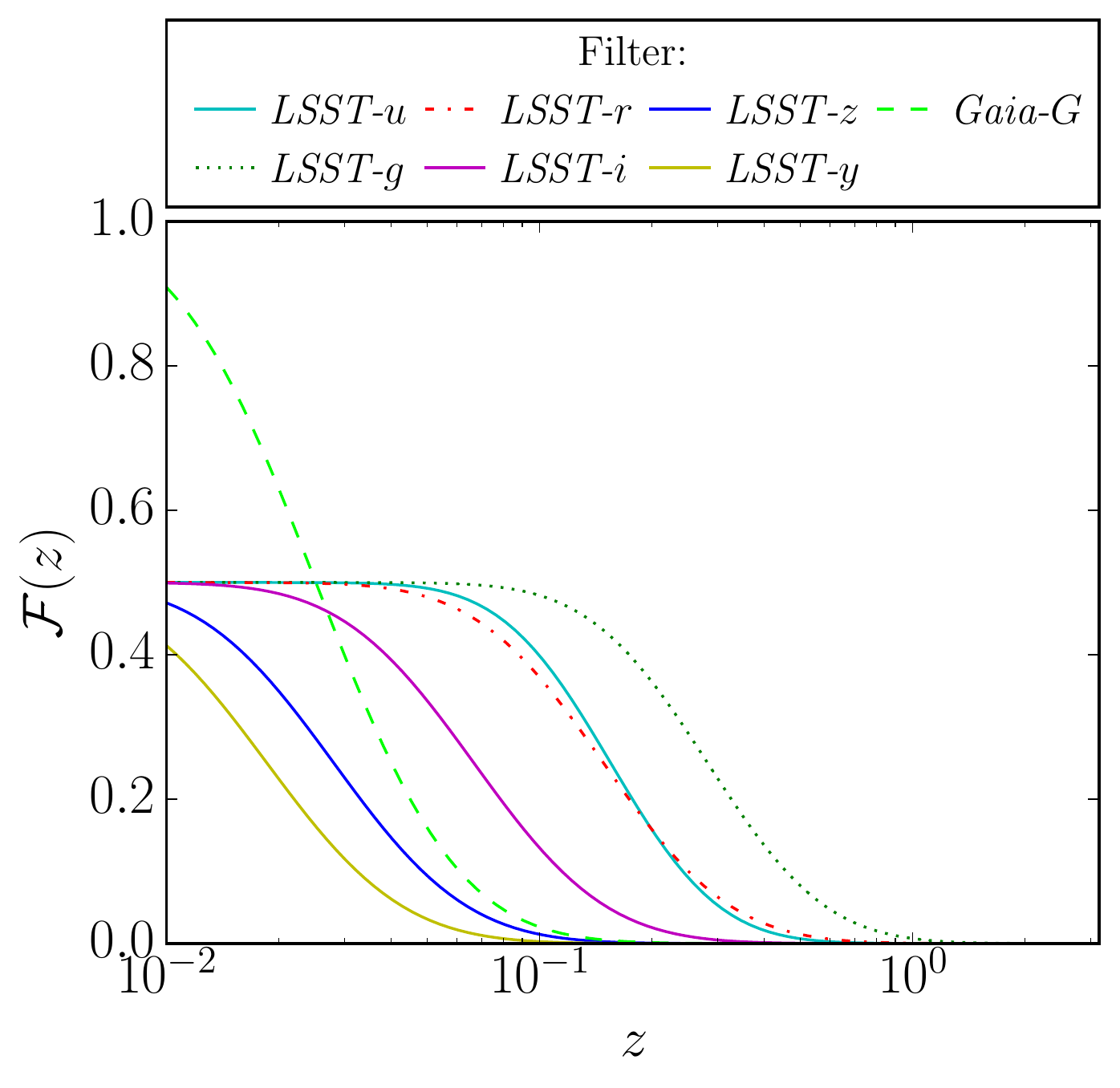}
	\caption{Completeness curves for \textit{LSST}'s and \textit{Gaia}'s filters, representing the fraction of global events at a given redshift which would be observable. It should be noted that the curves for \textit{LSST} are capped at 0.5, due to the survey only covering around half of the total sky.}
	\label{fig:observabilitycurves}
\end{figure}

Following the method described for \textit{LSST} in Section \ref{sec:detectors-LSST}, we calculate $z_\text{max}(M_{\rm BH})$ curves for \textit{Gaia}. From these, we derive fitting parameters to be used in equations \eqref{eq:fmz} and \eqref{eq:Fmz}, obtaining values of $B_G = -1.59967$ and $\sigma_G=0.30024$. It should be noted that when using equation \eqref{eq:Fmz} to calculate observability with \textit{Gaia}, we drop the extra factor of $1/2$ introduced when considering \textit{LSST}, since \textit{Gaia} covers the whole sky. This is evident in Fig.\ \ref{fig:observabilitycurves}, where we plot $\mathcal{F}_\phi(z)$ for \textit{LSST} and \textit{Gaia}.
\par
In order to calculate the detection rate of events in \textit{LSST} or \textit{Gaia}, we apply our completeness function, $\mathcal{F}_\phi(z)$, to the TDE rate rates derived in Section \ref{sec:tdrates} and integrate to obtain (for a filter, $\phi$)
\begin{multline}
	\frac{\dd{\mathcal{N}_{b,\phi}}}{\dd{t_0}} = \iiint\frac{\dd[4]{\mathcal{N}_{b,\text{tot}}}}{\dd{\log M_1}\,\dd{\log q}\,\dd{z}\,\dd{t_0}}\\\times\mathcal{F}_\phi(z)\,\dd{\log M_1}\,\dd{\log q}\,\dd{z},
	\label{eq:obsTDRateMBHB}
\end{multline}
for MBHBs, and
\begin{equation}
	\frac{\dd{\mathcal{N}_{s,\phi}}}{\dd{t_0}} = \iint\frac{\dd[3]{\mathcal{N}_{s,\text{tot}}}}{\dd{\log M_{\rm BH}}\,\dd{z}\,\dd{t_0}}\mathcal{F}_\phi(z)\,\dd{\log{M_{\rm BH}}}\,\dd{z},
	\label{eq:obsTDRateMBH}
\end{equation}
for single MBHs. As mentioned above, when considering \textit{Gaia} we multiply $\mathcal{F}_\phi(z)$ given in equation \eqref{eq:Fmz} by 2, to account for \textit{Gaia}'s full sky coverage.

\subsection{\textit{eROSITA}}
\label{sec:detectors-eROSITA}
The \textit{eROSITA} All-Sky Survey (eRASS) is an upcoming X-ray survey with the \textit{eROSITA} telescope \citep{erosita12}, due for launch in early 2019. The survey will cover the entire sky once every six months, for a four-year period. It will scan in the 0.5--2~keV band, so we focus on this band for our X-ray observability calculations. We follow a procedure similar to that employed for {\it LSST} and {\it Gaia}, with some instrument-specific variation, as described in the following.

From Section \ref{sec:emission}, we can generate a randomized light curve for an event sampled in a particular bin of the MBHB TDE distribution. The peak luminosity is found by integrating our phenomenological spectrum, and a decay time-scale is sampled from the $\tau$ column in Table \ref{tab:XrayTaus}.
The first measurement of the event will occur anywhere within the first six months after the event, with successive measurements at six-month intervals after that. This means the decay time-scale becomes comparable to the time over which the event is observed, and therefore this time-scale becomes a more significant factor in the observability of events. The time-scale must be long enough that the light curve remains observable for up to 1.5~years after the disruption event, but short enough that two magnitudes of decay can be observed in that period.

The \textit{eROSITA} Science Book \citep{erosita12} gives the flux sensitivity of \textit{eROSITA} for point sources. The value given for the average sensitivity after a single six-month period -- the most appropriate value for transient events -- is $4.4\e{-14}~\si{erg.s^{-1}.cm^{-1}}$ \citep[p.~23]{erosita12}.
Due to the proposed scanning path of \textit{eROSITA}, some regions of the sky will be visited more regularly than others. The number of visits that a point on the celestial sphere receives per six months is proportional to $1/\cos{\theta}$, where $\theta$ is its latitude \citep[inferred from][fig.~5.7.3]{erosita12}. This ranges from 1 for points around the equator, up to $~457/8$ for points close to the poles. We slightly alter this distribution to prevent the function from going to infinity at the poles, using
\begin{equation}
    N_{\text{visits}}(\theta)=\max\left(1,\frac{1}{\cos{\theta}+8/457}\right).
    \label{eq:N_visits}
\end{equation}
This assures a minimum of 1 visit per six months. Wherever a non-integer value of $N_\text{visits}$ is sampled, we randomly assign an integer value based on how close the sampled value is to the integers above and below it. If $N_\text{visits}=2.75$, for example, we assign it 3 visits with 75 per cent probability and 2 visits with 25 per cent probability.

To sample latitudes of events on the celestial sphere, we use the probability density function
\begin{equation}
    P(\theta)\dd{\theta}\propto\cos{\theta}\dd{\theta},
    \label{eq:latitudeProb}
\end{equation}
corresponding to a uniform sky distribution. If a point receives multiple visits in a six-month period, we expect them to be close together in time -- corresponding to scans of adjacent strips of sky that overlap at the point in question. As such, we treat multiple visits as a single, longer visit, since the length of time taken to complete these visits is small compared to typical X-ray decay time-scales and the six-month cycle of \textit{eROSITA}. The effect of a longer visit is to improve the flux sensitivity of the survey by a factor of $\sqrt{N_\text{visits}}$, making events closer to the poles slightly more observable than those around the equator.

To calculate a completeness function for \textit{eROSITA}, we use Monte--Carlo methods to simulate 500 events in each $z$ and $M_1$ bin. The fraction of accepted events in each bin yields the two-dimensional completeness function $\mathcal{F}(M_1,z)$. We couple this to the cosmic TDE rates appropriate for MBHs and MBHBs derived in Section \ref{sec:tdrates}, and calculate the TDE detection rates using equations \eqref{eq:obsTDRateMBHB} and \eqref{eq:obsTDRateMBH}, but replace $\mathcal{F}(z)$ with $\mathcal{F}(M_1,z)$ (the same function can be used in both equations, as its form has no dependence on whether the disrupting black hole is part of a binary). There is no need to multiply $\mathcal{F}(M_1,z)$ by 1/2 in this calculation, as the latitude sampling in equation \eqref{eq:latitudeProb} covers the whole sky.

\section{Results}
\label{sec:results}

\subsection{Detection Rates}
Detection rates for  \textit{LSST}, \textit{Gaia} and \textit{eROSITA} are computed as described in the previous section, by applying equations \eqref{eq:obsTDRateMBHB} and \eqref{eq:obsTDRateMBH}. All the relevant numbers, discussed below, are listed in Table \ref{tab:resultstable}.

\begin{figure*}
	\centering
	\begin{minipage}{0.5\textwidth}
		\centering
		\includegraphics[width=\textwidth]{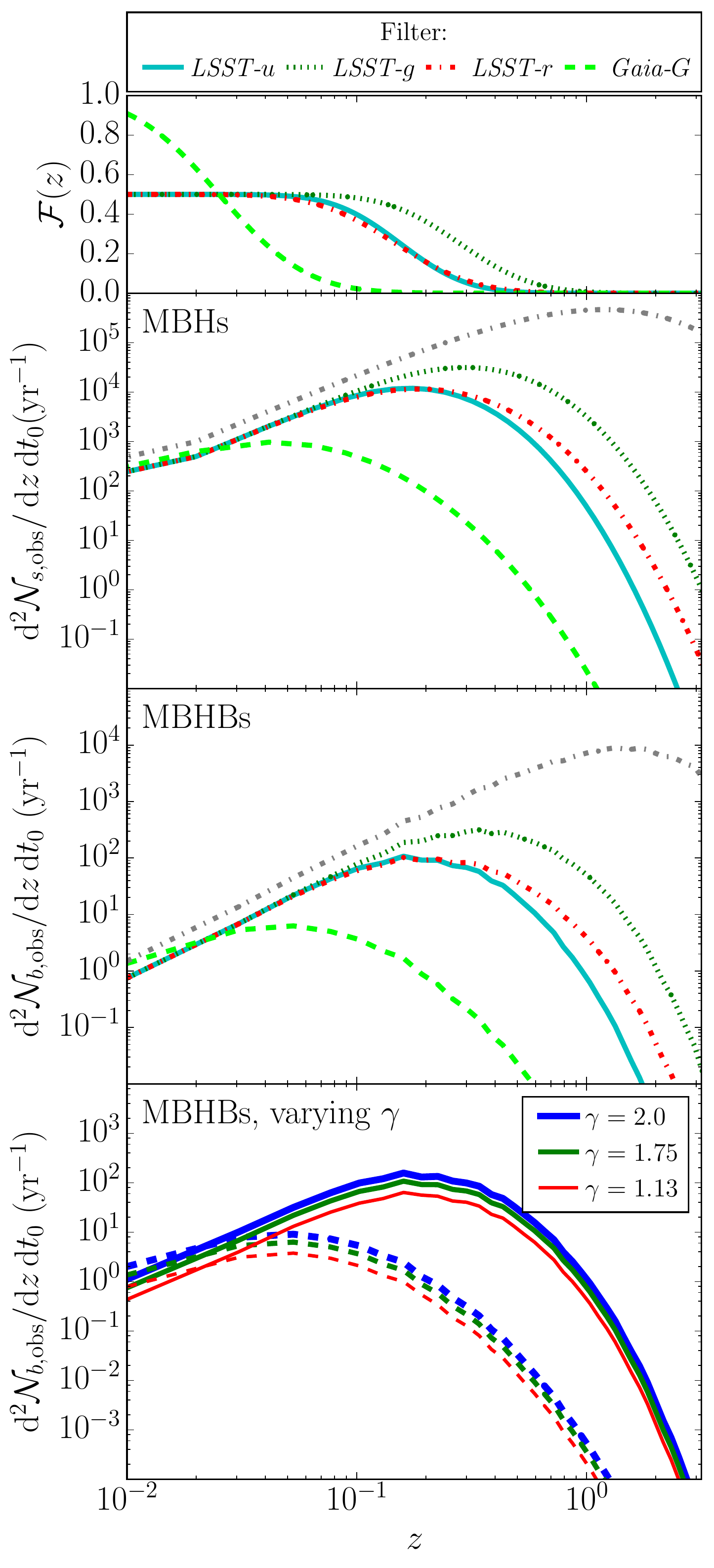}
	\end{minipage}%
	\begin{minipage}{0.5\textwidth}
		\centering
		\includegraphics[width=\textwidth]{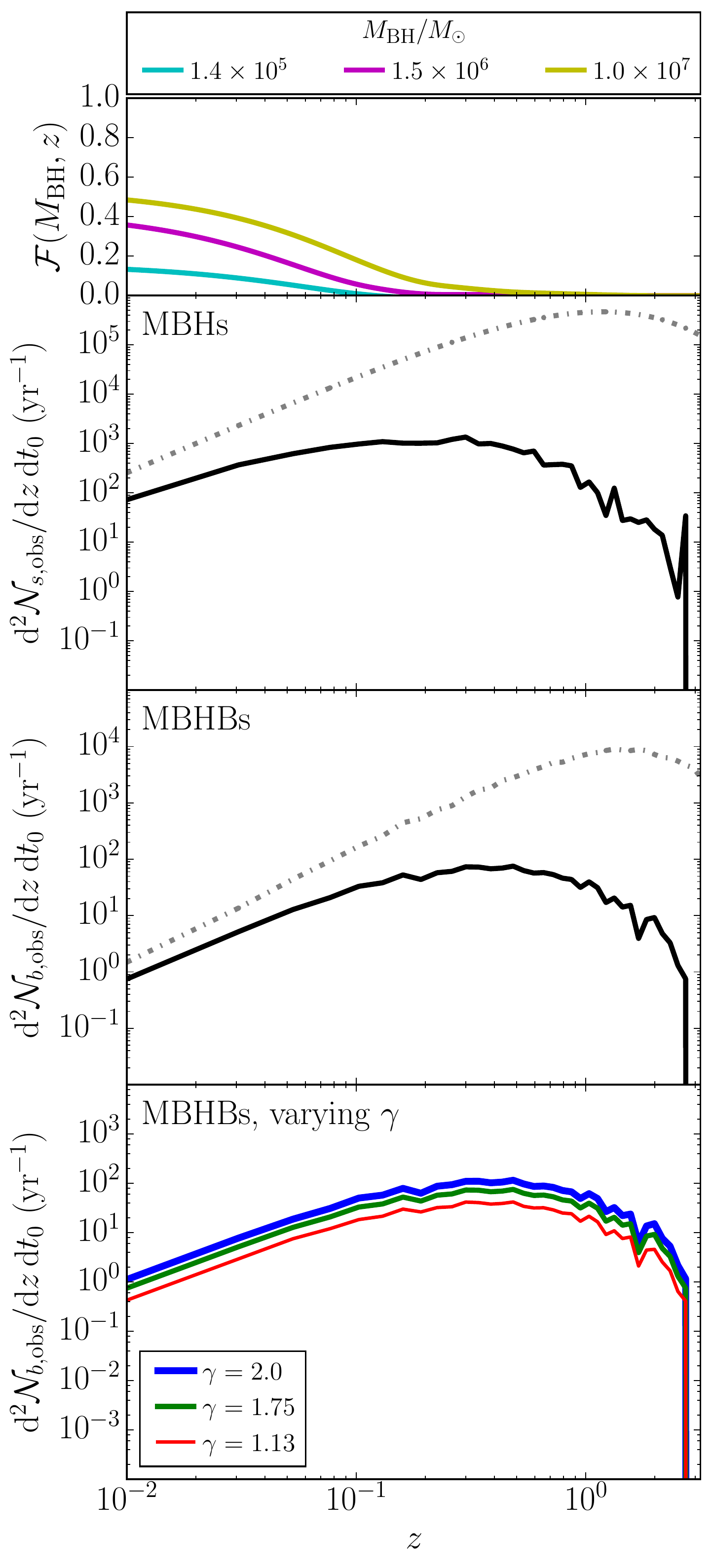}
	\end{minipage}
	\caption{TDE detection rates as a function of redshift. Left column: results for the optical surveys, \textit{LSST} and \textit{Gaia}. Here the top panel shows $\mathcal{F}$ for the three best {\it LSST} filters and for {\it Gaia}. The second and third panels down show the product of the TDE rate distributions in redshift with $\mathcal{F}$ in the different filters, for MBHs and MBHBs respectively ($\gamma=1.75$ is assumed). The thick dot-dashed grey line shows the global TDE rate for MBHs/MBHBs for comparison. The bottom panel shows the product of $\mathcal{F}$ with the MBHB TDE rate distribution for different values of $\gamma$. The solid lines in this plot represent \textit{LSST}-$u$, with dashed representing \textit{Gaia}. Right column: same results shown for \textit{eROSITA}. In this case, $\mathcal{F}$ is mass dependent and is shown for selected masses in the top panel. The second and third panels down again show the product of the TDE rate distributions in redshift with $\mathcal{F}$, for MBHs and MBHBs respectively (with $\gamma=1.75$) as thick solid black lines, with the thick dot-dashed grey line showing the global TDE rate for MBHs/MBHBs for comparison. Note that in this case solid lines are obtained using the mass-dependent $\mathcal{F}$. Finally, the bottom panel shows the product of $\mathcal{F}$ with the MBHB TDE rate distribution for different values of $\gamma$.} 
	\label{fig:stacks}
\end{figure*}

In the optical window, for \textit{LSST} and \textit{Gaia}, we calculate observed disruption rates due to single MBHs of 2807~yr$^{-1}$ and 81~yr$^{-1}$, respectively (for the more extreme mass loss case discussed in Section \ref{sec:tdrates}, these rise to 6310~yr$^{-1}$ and 182~yr$^{-1}$, respectively). For \textit{LSST}, we have considered detection in the $u$ filter, since we find this to be the third most successful at detecting TDEs, and want disruptions to be observed in three filters. When considering MBHBs, the result depends on the choice of $\gamma$ used when calculating the TDE rate distributions, as described in Section \ref{sec:populations-binarytdrate}. Assuming typical Bahcall--Wolf cusps of $\gamma=1.75$, we expect a detection rate due to binaries of around 31~yr$^{-1}$ for \textit{LSST}, meaning around 1.1 per cent of tidal disruptions detected by \textit{LSST} would be due to MBHBs (0.5 per cent if the higher limit of the single MBH TDE rate is used). For \textit{Gaia}, we find less than one event per year due to MBHBs, but could still expect up to 3 detections over the 5~yr mission. When carrying out these calculations, the result is affected by the detector's completeness function, $\mathcal{F}(z)$, and the TDE rate distribution. The impact of both of these factors is illustrated on the left-hand side of Fig.\ \ref{fig:stacks}. Here, the two central panels show the product of $\mathcal{F}(z)$ with the TDE rate distribution in redshift (with $\gamma=1.75$), for several different filters. The bottom panel shows the impact of varying $\gamma$, and thus the MBHB TDE rate distribution. Despite $\gamma$ having a mild effect on the global rate, it does not affect the redshift distribution of TDE detections. {\it LSST} is expected to observe one TDE per year out to $z\approx1$, with the detection rate broadly peaking at $0.1<z<0.5$. Conversely, {\it Gaia} might detect TDEs out to $z\lesssim0.2$, with a much shallower redshift dependence. 
\par
When considering \textit{eROSITA}, we
find detection rates of 669~yr$^{-1}$ and 71~yr$^{-1}$ from single MBHs and MBHBs respectively, where in the latter case, we have again assumed a Bahcall--Wolf cusp of $\gamma=1.75$. Assuming higher mass loss per disruption in the single MBH TDE rates causes the former value to rise to 849~yr$^{-1}$. This means that around 9.6 per cent of disruptions (7.7 per cent when higher mass loss is assumed) detected by \textit{eROSITA} are likely to be from binaries -- a much higher fraction than expected for \textit{LSST}. The cause for this lies in the fact that the X-ray event luminosity predicted by our phenomenological model tends to be correlated with mass more strongly than optical luminosity, which is largely mass independent (see Fig.\ \ref{fig:bandLums}). This means that \textit{eROSITA} will favour detection of events coming from the more massive end of the MBHB population, where the disruption rate tends to be higher (see left panel of Fig.\ \ref{fig:dNdlogTDE}). This effect is further emphasized by the slightly lower detection rate of TDEs from lone MBHs found for \textit{eROSITA}, leading to a higher binary fraction for \textit{eROSITA}. The impact of the X-ray luminosity's mass dependence on the effective completeness function, $\mathcal{F}(M_{\rm BH},z)$, for \textit{eROSITA} is shown in the upper panel on the right hand side of Fig.\ \ref{fig:stacks}. The redshift distributions shown in the panels below highlight the higher potential of \textit{eROSITA} of detecting TDEs from MBHBs at a rate of one per year out to $z>2$. The redshift dependence of the detection rate is similar to that seen in \textit{LSST}, featuring a broad plateau in the range $0.2<z<1$.
\par

\begin{figure}
	\centering
	\includegraphics[width=\linewidth]{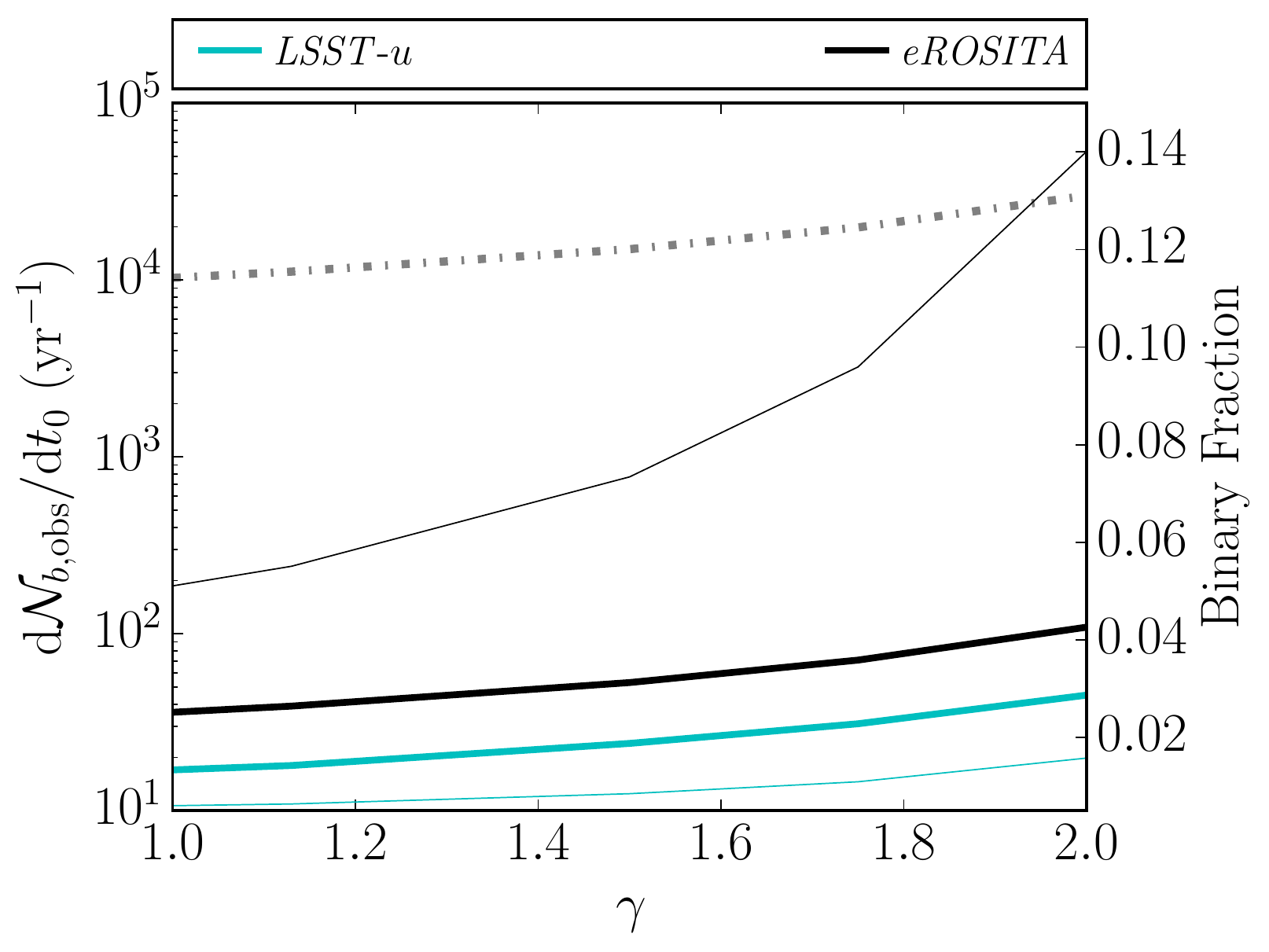}
	\caption{Expected detection rates of binary induced TDEs for \textit{LSST} and \textit{eROSITA} as a function of $\gamma$. The thick lines apply to the left hand axis, and show the numbers of detections per year. The dot-dashed line shows the global MBHB TDE rate, as a point of comparison. The thin lines show the fraction of detected TDEs which will be made up by those coming from binaries, and apply to the right hand axis.}
	\label{fig:tdrateslines}
\end{figure}

Fig.\ \ref{fig:tdrateslines} shows the MBHB TDE detection rate as a function of $\gamma$ for \textit{LSST} and \textit{eROSITA}. This confirms that $\gamma$ has a mild influence (less than a factor of three) on the detection rate for both instruments. It also highlights the larger fraction of total TDEs due to MBHBs for {\it eROSITA}. As previously discussed, the X-ray luminosity of TDEs appears to correlate linearly with MBH mass, whereas the optical luminosity does not. Since regular TDEs are hosted preferentially by low mass MBHs and MBHB TDEs preferentially occur in high mass systems, a flux-limited X-ray survey will naturally select the latter.
\par
\begin{figure}
	\centering
	\includegraphics[width=\linewidth]{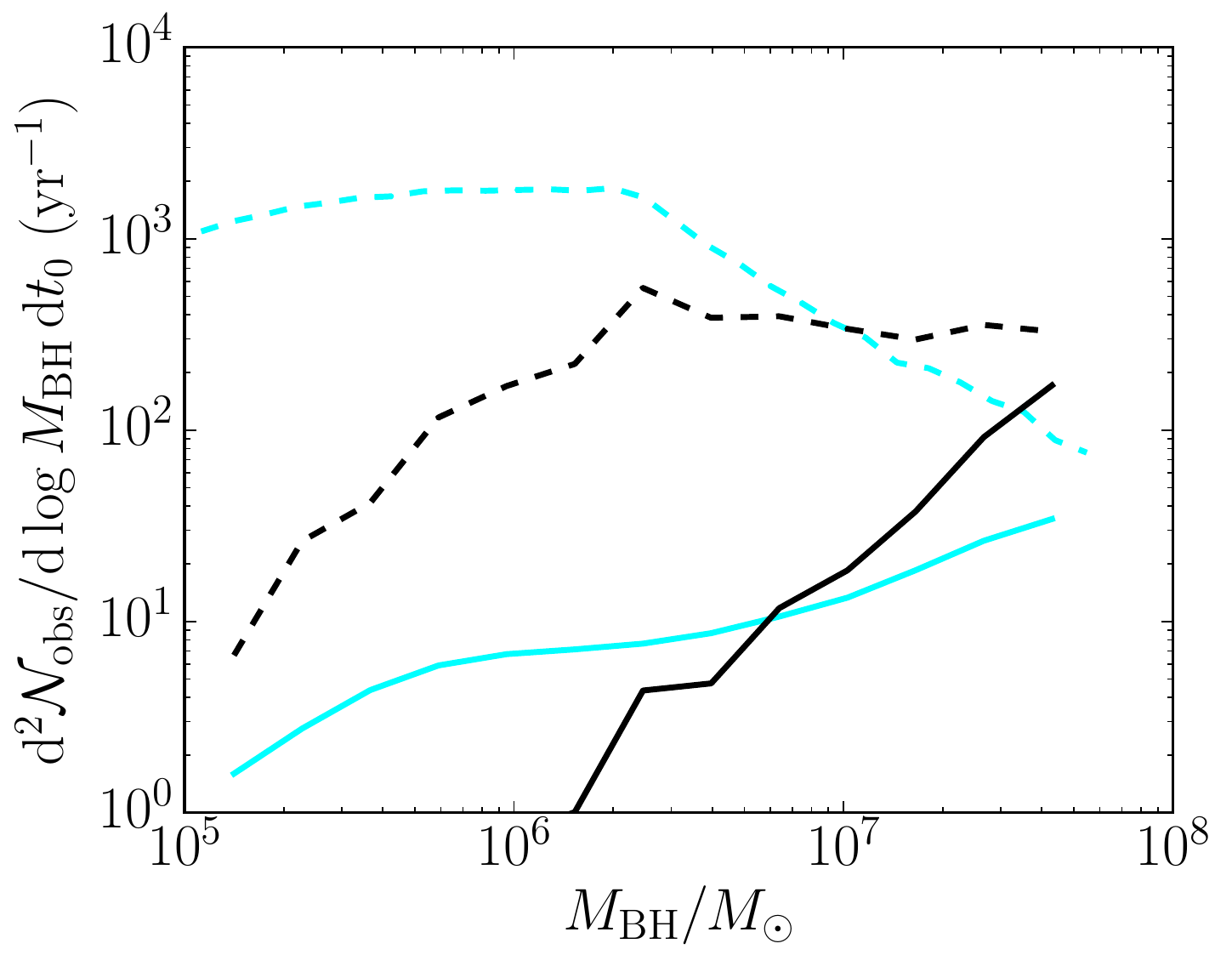}
	\caption{Mass functions of the TDEs observed by \textit{LSST} (cyan lines) and \textit{eROSITA} (black lines). Dashed lines are single MBHs, whilst solid lines are disruptions from MBHBs. In the binary case, primary black hole mass is being considered, and a cusp with $\gamma=1.75$ is assumed.}
	\label{fig:obsmassfns}
\end{figure}
The mass functions of TDEs detected by \textit{LSST} and \textit{eROSITA} are plotted in Fig.\ \ref{fig:obsmassfns}. For \textit{LSST}, the observed mass functions approximately follow the trend of the underlying differential TDE rates calculated in Section \ref{sec:populations} (c.f.\ Fig. \ref{fig:massfuncs} and the left panel of Fig.\ \ref{fig:dNdlogTDE}). For \textit{eROSITA}, the trend is somewhat similar, but with a much more aggressive drop off with black hole mass. This is due to \textit{eROSITA}'s preference for more massive black holes, as already discussed above. One may have hoped that a difference in the TDE mass functions of single MBHs and MBHBs could act as a signpost to the existence of MBHBs. Whilst the components of the overall TDE mass function will be different, the MBHB TDE rate appears to be sufficiently sub-dominant that any signature of this difference will likely not be visible in the overall observed mass function. Having said that, Figs.\ \ref{fig:massfuncs} and \ref{fig:obsmassfns} suggest that TDEs around the most massive black holes (i.e. with $M_{\text{BH}}\gtrsim 3\times 10^7M_\odot$) are more likely to have a binary origin. However, even at the highest masses, the rates are still dominated by single MBHs, and distinctive signatures to identify MBHBs should be sought. 

\begin{figure}
	\centering
	\includegraphics[width=\linewidth]{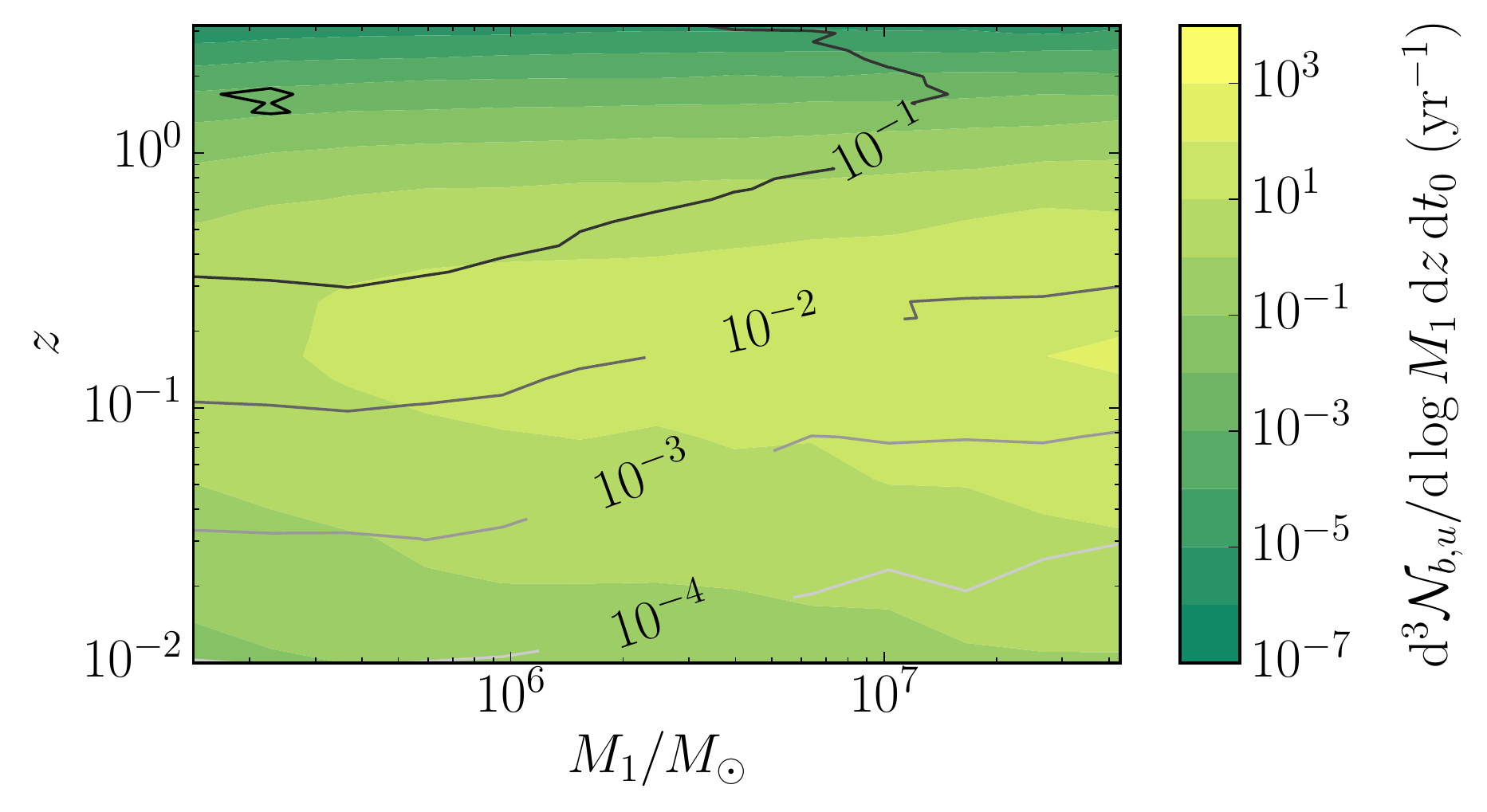}
	\includegraphics[width=\linewidth]{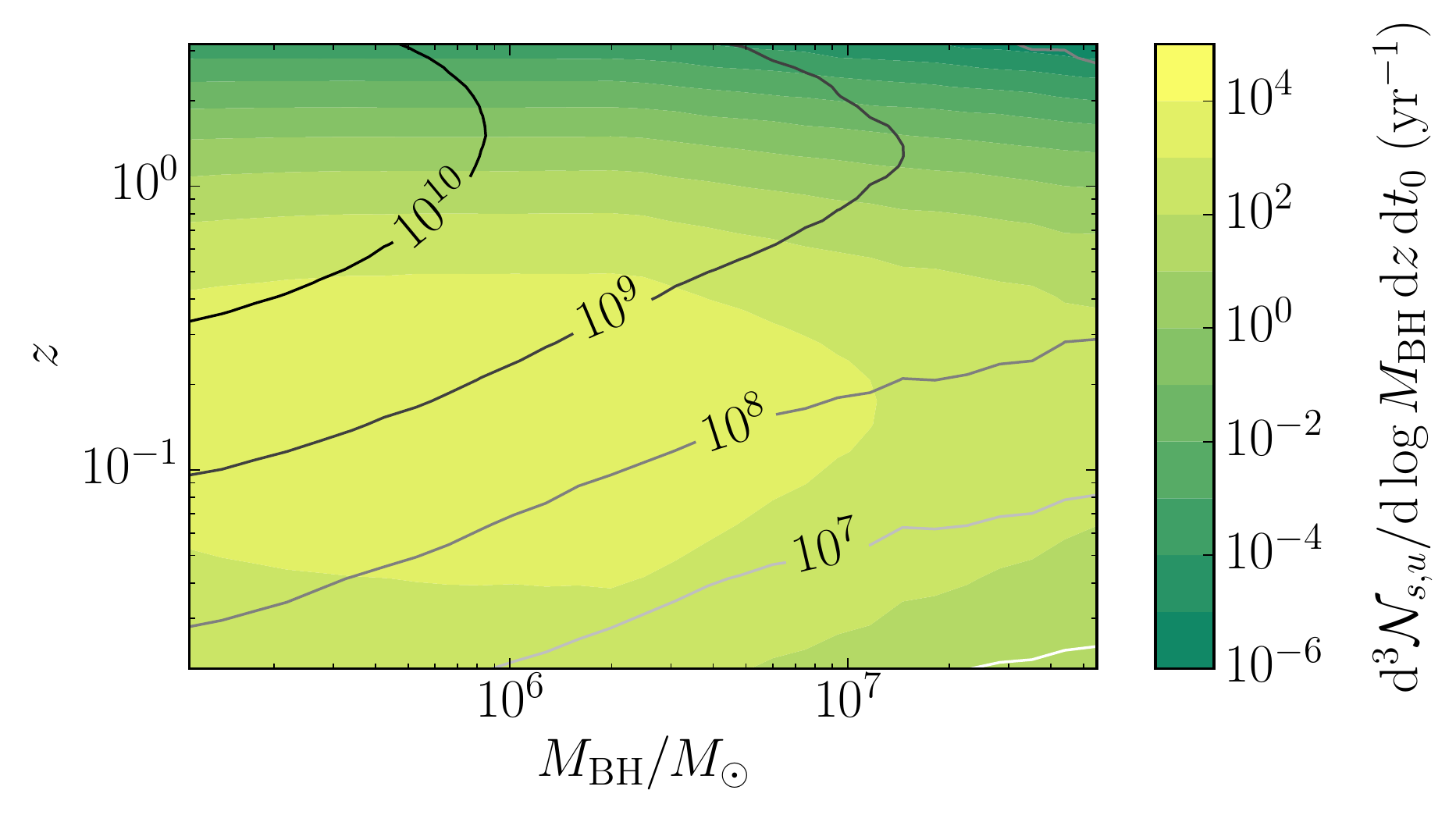}
	\caption{Top panel: contour plots contrasting the observed TDE rate from MBHBs -- $\dd[3]{\mathcal{N}_{b,\text{obs}}}/(\dd{\log {M_1}}\,\dd{z}\,\dd{t_0})$, filled contours -- and the underlying MBHB merger rate -- $\dd[3]{N_{b}}/(\dd{\log {M_1}}\,\dd{z}\,\dd{t_0})$, empty contours -- from the Millennium run. Bottom panel: contour plots contrasting the observed TDE rate from lone MBHs -- $\dd[3]{\mathcal{N}_{s,\text{obs}}}/(\dd{\log {M_{\rm BH}}}\,\dd{z}\,\dd{t_0})$, filled contours --  and the underlying MBH population -- $\dd[2]{N_s}/(\dd{\log{M_{\text{BH}}}}\,\dd{z})$, empty contours -- from the Millennium run. In both cases we considered TDEs detected by {\it LSST}.}.
	\label{fig:MBHcontours}
\end{figure}

Finally, it is interesting to consider how representative observed TDEs triggered by MBHBs (or lone galactic centre MBHs) are of the underlying MBHB (or lone MBH) population in the Universe. This is shown for \textit{LSST} detections, in Fig.\ \ref{fig:MBHcontours}. In the top panel, the observed MBHB TDE rate distribution is compared to the MBHB merger rate distribution derived from Millennium--II. We can see that the population of MBHBs likely to be revealed through TDE observations is not representative of the global MBHB population, being heavily weighted towards much higher primary masses. The effect is even more pronounced for \textit{eROSITA} (not shown), because of the aforementioned dependence of TDE luminosity on MBH mass. Conversely, the bottom panel shows that the lone MBH population revealed through TDE observation is more representative of  (although shallower than) the underlying MBH mass distribution.

Incidentally, the total number of expected MBHB TDE detections over the planned survey lifetime is similar for \textit{LSST} and \textit{eROSITA}, spanning the range 150--450 depending on $\gamma$, as reported in Table \ref{tab:resultstable}.

\begin{table*}
	\centering
	\caption{Numbers of binary induced disruptions observed over the survey duration of each of the detectors considered, for different assumed values of $\gamma$. Also included is the number of binaries we expect to observe disrupting exactly twice, or at least thrice. The second-last column lists the fraction of all observed disruptions we expect to be due to binaries, and the final column lists the fraction of double flares which will be observed from single black holes (i.e. the fraction of contamination we expect in a sample of binaries inferred from double flares). In these two rightmost columns, the bracketed values are calculated assuming an MBH mass growth per disruption of $0.15M_\odot$, rather than the default of $0.45M_\odot$ (see Section \ref{sec:tdrates} for more detail).}
	\label{tab:resultstable}
	\begin{tabular}{c c c c c c c}\toprule
		Detector & $\gamma$ & $\mathcal{N}_{b,\text{obs}}$ & $N_b(n=2)$ & $N_b(n\geq3)$ & $\mathcal{N}_{b,\text{obs}}/\mathcal{N}_{b+s,\text{obs}}$~(\%) & $N_s(n=2)/N_{b+s}(n=2)$~(\%)\\\midrule
		\textit{LSST}-$u$ & 2.0 & 448 & 31 & 53 & 1.58 (0.71) & 17.84 (44.44)\\
			& 1.75 & 306 & 33 & 12 & 1.09 (0.49) & 16.85 (42.76)\\
			& 1.5 & 236 & 16 & 4 & 0.85 (0.38) & 29.87 (61.08)\\
			& 1.13 & 180 & 7 & 1 & 0.64 (0.28) & 49.37 (78.23)\\
		\textit{eROSITA} & 2.0 & 438 & 61 & 43 & 14.06 (11.41) & 0.49 (0.83)\\
			& 1.75 & 283 & 23 & 3 & 9.58 (7.70) & 1.26 (2.13)\\
			& 1.5 & 211 & 8 & $1$ & 7.30 (5.84) & 3.42 (5.67)\\
			& 1.13 & 156 & 3 & $<1$ & 5.49 (4.37) & 8.72 (13.96)\\
		\textit{Gaia}-$G$ & 2.0 & 5 & $<1$ & $<1$ & 1.23 (0.55)& -- \\
			& 1.75 & 3 & $<1$ & $<1$ & 0.84 (0.38)& -- \\
			& 1.5 & 3 & $<1$ & $<1$ & 0.65 (0.29)& -- \\
			& 1.13 & 2 & $<1$ & $<1$ & 0.50 (0.22) & -- \\
		\bottomrule
	\end{tabular}
\end{table*}

\subsection{Distinguishing MBHBs TDEs from Single MBHs TDEs}
\subsubsection{Recurrent Flares}
\label{sec:distinction}
A mentioned in Section \ref{sec:introduction}, the high rates of disruption thought to be possible from MBHBs could lead to multiple disruptions from the same binary being visible over the course of one of the surveys we are considering. Identifying such occurrences could be an effective way of probing for galactic centre binaries throughout the universe.
\par
In order to predict the number of recurrent flares which could be observable by a particular survey, we define the expected number of disruptions, $\mu$, from an MBH or MBHB which disrupts stars at a rate $\dot{\mathcal{N}}$ as being
\begin{equation}
	\mu = \frac{\dot{\mathcal{N}}}{1+z}\times t_\text{survey}.
\end{equation}
Here, $t_\text{survey}$ is the duration of the survey being considered and $\dot{\mathcal{N}}$ equals either $\dot{\mathcal{N}}_{s}(M_{\rm BH})$ or $\dot{\mathcal{N}}_{b}(M_1,q,\gamma)$, given by equations \eqref{eq:ntdsinglee} and \eqref{eq:ntd}, respectively. Then, by Poisson statistics, the probability the MBH/MBHB disrupts $n$ stars over a time $t_\text{survey}$ will be given by
\begin{equation}
	P(n|\mu) = \frac{\mu^ne^{-\mu}}{n!},
\end{equation}
from which it follows that the probability of at least some critical number, $n_\star$, of disruptions will be
\begin{equation}
	P(n\geq n_\star|\mu) = 1-\sum_{n'=0}^{n_\star-1}\frac{\mu^{n'}e^{-\mu}}{n'!}.
\end{equation}
\par
In order to apply this to our MBHB population, we must account for the fact that only those MBHBs which are in the enhanced disruption rate phase of their lifetime will be disrupting at a rate of $\dot{\mathcal{N}}_{b}(M_1,q,\gamma)$. We find a distribution of enhanced phase binaries by multiplying our binary merger rate distribution by $t_D(M_1,q,\gamma)$, given by equation \eqref{eq:tdutycycle}:
\begin{multline}
	\frac{\dd[3]{N_{b,\text{enhanced}}}}{\dd{\log M_1}\,\dd{\log q}\,\dd{z}} \\= (1+z)t_D\times\frac{\dd[4]{N_b}}{\dd{\log M_1}\,\dd{\log q}\,\dd{z}\,\dd{t_0}}.
\end{multline}
\par
Calculating the number of MBHs or MBHBs which will produce at least $n_\star$ observable disruptions over a survey's lifetime is then achieved using
\begin{multline}
	N_b(n\geq n_\star) = \iiint\frac{\dd[3]{N_{b,\text{enhanced}}}}{\dd{\log M_1}\,\dd{\log q}\,\dd{z}}\\\times\mathcal{F}\times P(n\geq n_\star|\mu)\,\dd{\log M_1}\,\dd{\log q}\,\dd{z},
\end{multline}
for an MBHB, or
\begin{multline}
	N_s(n\geq n_\star) = \iint\frac{\dd[2]{N_s}}{\dd{\log M_{\rm BH}}\,\dd{z}}\\\times\mathcal{F}\times P(n\geq n_\star|\mu)\,\dd{\log M_{\rm BH}}\,\dd{z},
\end{multline}
for a single MBH. Here, $\mathcal{F}$ is the completeness function for a particular instrument. Using $P(n|\mu)$ in place of $P(n\geq n_\star|\mu)$ in these integrals allows calculation of the number of MBHs or MBHBs which will produce exactly $n$ disruptions.
\par
\begin{figure}
	\centering
	\includegraphics[width=\linewidth]{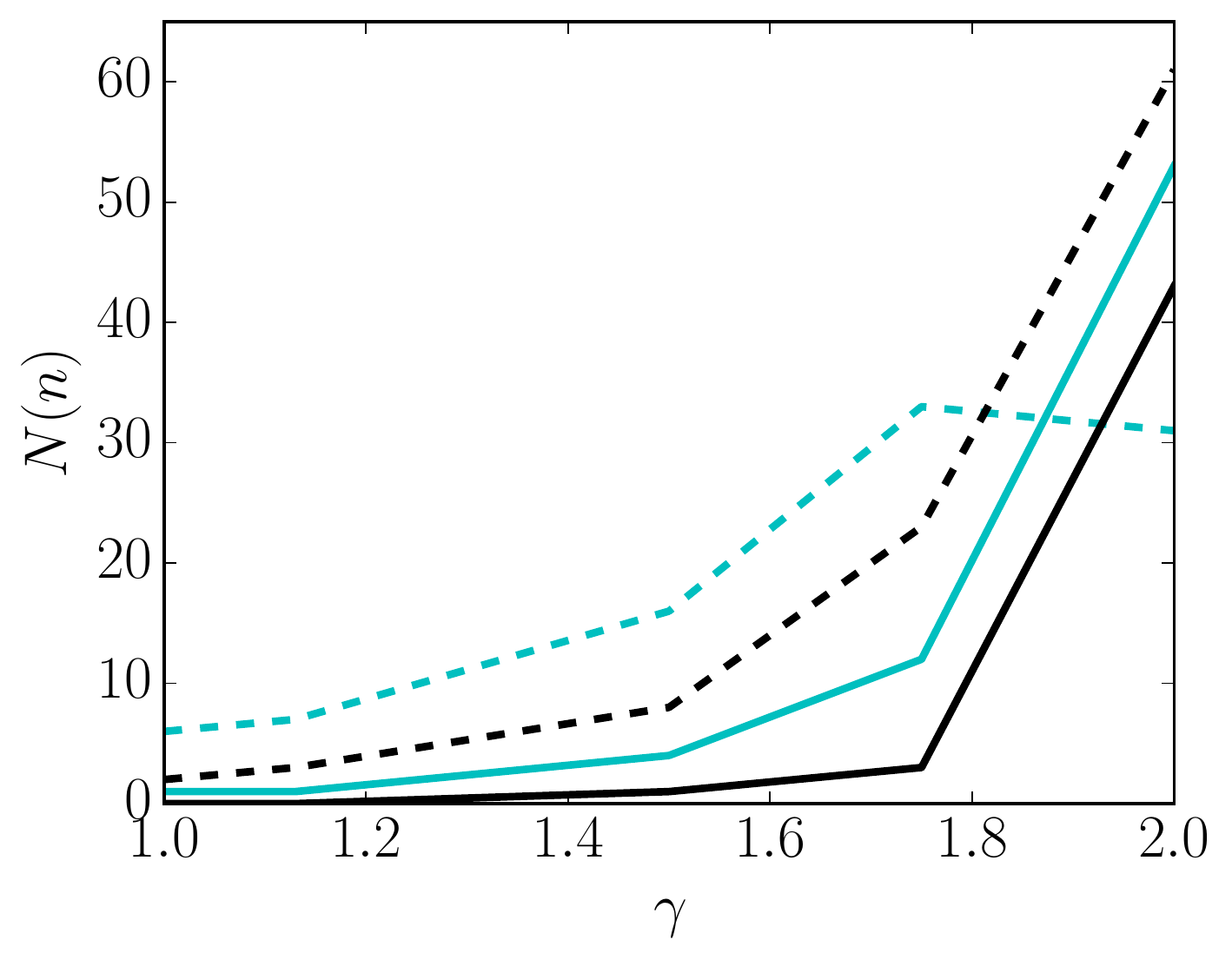}
	\caption{Numbers of MBHBs observed to produce multiple disruptions over the lifetimes of \textit{LSST} and \textit{eROSITA} (cyan lines indicate \textit{LSST}, black lines \textit{eROSITA}). Dashed lines show $N_b(n=2)$ -- i.e.\ the number of MBHBs producing exactly 2 disruptions. Solid lines show $N_b(n\geq3)$ -- i.e.\ the number producing three or more disruptions.}
	\label{fig:recurrentlines}
\end{figure}
\par
We use this to calculate the number of MBHBs which will be seen to disrupt two stars over each survey's lifetime, i.e.\ $N_b(n=2)$. For our standard case of binaries embedded in a Bachall--Wolf cusp ($\gamma=1.75$), we find that over \textit{LSST}'s 10 year surveying duration it is likely to see around 33 MBHBs producing exactly two disruptions. Despite the shorter survey duration of 4 years for \textit{eROSITA}, we still expect to observe a similar number of double disruptions from MBHBs -- around 23 over its lifetime. Conversely, \textit{Gaia} is unlikely to observe any.
\par
To ascertain whether a double disruption can be taken as good evidence for the presence of an MBHB (as opposed to a single MBH) in a galaxy, we also calculate the number of single MBHs we expect each survey to observe producing two disruptions, i.e.\ $N_s(n=2)$. For \textit{LSST}, we expect to see 7 MBHs producing two disruptions over the survey duration (rising to 25 if the TDE rates are calculated for the more extreme mass loss case), whilst for \textit{eROSITA}, we are unlikely to see more than one MBH doing the same. This means that a sample of galactic centre MBHBs inferred from the observation of double disruptions with \textit{LSST} would be subject to a significant amount of contamination from single MBHs. Still, recurrent flares are an excellent means of identifying MBHBs.
\par
Classifying a galaxy as containing an MBHB only when it is seen to produce three disruptions over a survey's lifetime results in a far `purer' sample. Looking again at the case of $\gamma=1.75$, we find the number of MBHBs producing three or more disruptions, i.e.\ $N_b(n\geq3)$, to be 12 in the case of \textit{LSST}, and 3 in the case of \textit{eROSITA}. For both surveys, we expect no single MBHs to be seen to produce more than two disruptions. As such, galaxies observed as producing three or more TDEs by either \textit{LSST} or \textit{eROSITA} will be strong MBHB host candidates. Table \ref{tab:resultstable} lists in its third and fourth columns our calculations of $N_b(n=2)$ and $N_b(n\geq3)$ for different values of $\gamma$. Its final column lists the level of contamination we expect an MBHB sample formed of galaxies producing two TDEs to be subject to. Fig.\ \ref{fig:recurrentlines} shows the data graphically.

Several interesting points emerge from this. The first is that in all cases, \textit{eROSITA} suffers from less contamination of its double disruption sample by single MBHs. This is likely because the short survey duration of 4 years impacts the detection of multiple disruptions from single MBHs more than it does than it does those from MBHBs. A second feature of note is that for $\gamma>1.75$, the number of MBHBs observed by \textit{LSST} as disrupting twice over the 10 years actually starts to decrease again (seen most clearly in Fig.\ \ref{fig:recurrentlines}). This is due to the fact that the very high disruption rates expected as $\gamma\to2$ mean that a large fraction of the MBHBs which disrupt twice over \textit{LSST}'s lifetime will be likely to disrupt one or more further times. This gives a very high value of $N_b(n\geq3)$, at the expense of the value of $N_b(n=2)$. Such behaviour is not seen for \textit{eROSITA}, as the short survey duration makes a triple disruption more difficult, even for an extreme value of $\gamma$.

\begin{figure}
	\centering
	\includegraphics[width=\linewidth]{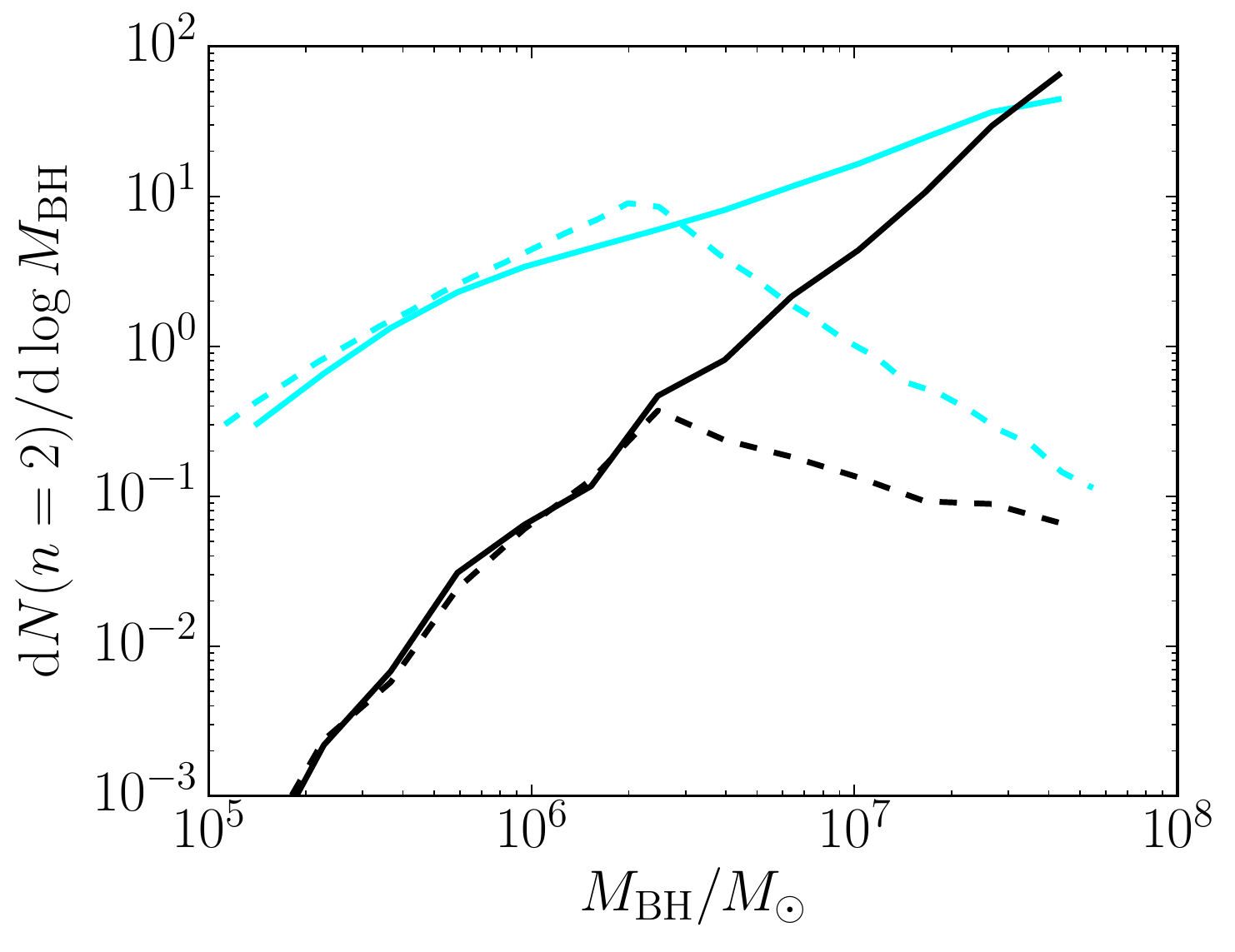}
	\caption{Mass functions of MBHs observed as disrupting twice by \textit{LSST} (cyan lines) and \textit{eROSITA} (black lines) during their respective mission lifetimes. Solid lines are for TDEs induced by MBHBs, whereas dashed lines are for standard TDEs from single MBHs. In the MBHB case, the mass of the primary hole is considered and $\gamma=1.75$ is assumed.}
	\label{fig:recmassfns}
\end{figure}

One final thing about the population of recurrently flaring MBHs and MBHBs which we can compare is their mass functions. Fig. \ref{fig:recmassfns} shows the mass functions of MBHs and MBHBs disrupting twice during the \textit{LSST} and \textit{eROSITA} lifetimes. The two detectors find mass functions of fairly similar shapes, albeit with a steeper mass dependence in the results for \textit{eROSITA} (as also seen in Fig.\ \ref{fig:obsmassfns}). Since, the TDE rate from single MBHs is a declining function of mass for $M_{\text{BH}}>10^6\msun$, it is extremely unlikely to have recurrent flares from single MBHs with $M_{\text{BH}}>10^7\msun$ (in fact, less than one such occurrence is expected both for \textit{LSST} and \textit{eROSITA}). Therefore, any system with inferred disrupting MBH mass $>10^7\msun$ showing recurrent flares can be considered a very strong MBHB candidate.

\subsubsection{Interrupted flares}
Another possible signature of MBHB tidal disruption has been identified in temporary interruptions to the supply of fallback material due to the secondary black hole intersecting and perturbing the returning gas stream, as originally proposed by \cite{liu09}. The prediction has been corroborated by a compilation of SPH simulations performed by \cite{coughlin17}, that showed that the perturbing effect of the secondary results in a complex phenomenology. This includes distinct dips in the measured mass fallback, although more erratic deviations from the expected $t^{-5/3}$ return time distribution are generally observed. The optical lightcurve of the inactive galaxy SDSS J120136.02+300305.5 shows dips on roughly a month time-scale, consistent with those predicted theoretically, and has been put forward as a TDE candidate from an MBHB with $M_1=10^7\msun$, $q=0.8$ and binary separation $a=0.6$~mpc \citep{2014ApJ...786..103L}. 

Although interrupted flares are an appealing feature of MBHB TDEs, it should be noticed that such signature typically occurs for extremely compact, sub-mpc binaries (as those considered in the aforementioned theoretical works), which are likely in the hard binary phase and have already disrupted their cusp of bound stars \citep{2010ApJ...719..851S}. They are therefore beyond the binding phase which leads to the short burst of TDEs modelled by \cite{chen11} and considered in this work, as we now show.

Starting from \cite{liu09}, the typical time-scale of flare interruption is given as
\begin{equation}
  T_{\rm int}\approx \frac{P_b}{4.7}(1+e)^{-3/5}(1-e)^{9/5}\left(1-\frac{0.3\theta}{\pi}\right)^{-3/2},
  \label{eq:treturn}
\end{equation}
where $P_b$ is the binary orbital period, $e$ is its eccentricity, and $\theta$ is the inclination between the MBHB and the disrupted stars' orbital plane. Here and in the following we ignore $(1+q)$ factors, considering binaries with $q\ll 1$. The cusp erosion phase modelled in this work starts at a separation $a_0$ for which $M_*(a<a_0)\approx M_2$, where $M_*(a<a_0)$ is the stellar mass contained within a sphere of radius $a_0$ around the primary black hole. For the cuspy stellar distribution considered here, $a_0$ is of the order \citep{chen11}
\begin{equation}
  a_0\approx (3-\gamma)\pfracp{M_1}{10^6\msun}{1/2}q^{1/(3-\gamma)}~\text{pc}.
  \label{eq:a0}
\end{equation}
This phase lasts roughly until the MBHB has shrunk to $a\approx a_0/10$. Inserting equation \eqref{eq:a0} into equation \eqref{eq:treturn} we get
\begin{multline}
  T_{\rm int}\approx 1.9\times10^4\,\,(3-\gamma)^{3/2}\pfrac{M_1}{10^6\msun}\,q^{3(3-\gamma)/2} \\
  \times\left(\frac{a}{a_0}\right)^{3/2}(1+e)^{-3/5}(1-e)^{9/5}~\text{yr}.
  \label{eq:treturn2}
\end{multline}
Although apparently hopeless, it should be noticed that for $e=0$, $\gamma=2$, $q=0.01$ and $a/a_0=0.1$, the above equation yields $T_{\rm int}\approx 0.6$~yr for $M_1=10^6\msun$, reducing to $T_{\rm int}\approx 20$~d for $M_1=10^5\msun$. In fact, these binaries have a separation $a=1$~mpc and $a=0.3$~mpc, respectively, and interrupted flares are therefore expected. Further, we note that these time-scales can be vastly shortened for large values of $e$. If eccentric unequal binaries are common, as suggested by hybrid evolution models \citep[e.g.][]{2010ApJ...719..851S} as well as numerical simulations \citep[e.g.][]{2007ApJ...656..879M}, then interrupted flares will also be common among less compact binaries. For $e=0.9$, the normalization of equation \eqref{eq:treturn2} drops by nearly two orders of magnitude, to $\sim 300$\, yr. In this case binaries with $q \approx 0.1$ residing in shallower cusps will also experience flare interruptions on month time-scales, making this signature common among our systems. Since the occurrence of interrupted flares is highly dependent on the unknown MBHB eccentricity, we do not attempt to make predictions here. We just notice that the {\it absence} of flare interruptions is not sufficient to discard the presence of a sub-pc MBHB.

\section{Discussion and conclusions}
\label{sec:discussion}
In this work we performed detailed predictions for the expected cosmic rates of MBH and MBHB TDEs -- with a specific focus on the latter -- in an attempt to quantify the potential of TDEs to probe the cosmic population of sub-pc MBHBs. To this end, we combined MBH and MBHB populations derived from the \texttt{guo2010a} semianalytic galaxy formation model applied to the high resolution cosmological simulation Millennium--II, with estimates of the induced TDE rates for each class of objects. For individual MBHs, we used the TDE rates empirically derived by \cite{stone16}, whereas for sub-pc MBHBs we relied on the analytical calculations of \cite{chen11}, built upon detailed three-body scattering experiments. We then constructed empirical TDE spectra that fit a large number of observations in the optical, UV and X-ray, and considered their observability by current and future survey instruments, including \textit{LSST} and {\it Gaia} in optical and \textit{eROSITA} in X-ray. 
\par
Overall, we find a promising outlook for the detection of tidal disruption events with \textit{LSST} and \textit{eROSITA}, including the possibility of detecting a hitherto unseen number of MBHB induced events. Considering conventional tidal disruptions arising from single galactic centre MBHs, we expect 2807--6310 observations per year with \textit{LSST}, in reasonable agreement with the calculations of previous authors \citep[who predict rates of $6000~\si{yr}^{-1}$, $4180~\si{yr}^{-1}$ and $5003\pm1421~\si{yr}^{-1}$, respectively]{strubbe09,vanvelzen11,mageshwaranmangalam}. For \textit{eROSITA} we predict between 669 and 849 detections annually, which is again comparable to the results of \citet{mageshwaranmangalam}, who find an expected detection rate of $679.5\pm195$~yr$^{-1}$, and is also comparable to the results of \citet{khabibullin14}, who predict several thousand detections over the full length of eRASS. For \textit{Gaia}, our expected detection rates of 81--182~yr$^{-1}$ are significantly higher than that predicted by \citet{blagorodnova16}, who expect only 20--30 TDEs per year. This discrepancy becomes even more concerning when one considers that \textit{Gaia} has not yet alerted the detection of a single TDE. However, the recent work of \citet{kostrzewa18} uses a different transient selection method to the standard \textit{Gaia} Science Alerts team, and finds 160 nuclear transients in a year's worth of data which covers 1/3 of the sky. Scaling this up to the whole sky, one could expect a nuclear transient detection rate of $\sim480$~yr$^{-1}$, around 2.2 times greater than the 215~yr$^{-1}$ which \citet{kostrzewa18} quote \citet{blagorodnova16} as predicting. Whilst \citeauthor{kostrzewa18} do not attempt to classify candidate TDEs in their study, one could assume that the TDEs make up the same fraction of nuclear transients as predicted by \citeauthor{blagorodnova16}, meaning one can expect around 44--66 TDEs per year in the \textit{Gaia} data. Our result is still an overestimate compared to this, but the difference is slightly less substantial.
\par
Turning to the detection of MBHB induced TDEs, we predict sizeable numbers of events to be observed by both \textit{LSST} and \textit{eROSITA} (31~yr$^{-1}$ for the former, and 71~yr$^{-1}$for the latter). Our calculated rate for \textit{LSST} is a few times higher than that found by \citet{weggbode11}, who find a more conservative detection rate of 8~yr$^{-1}$. Previous predictions for \textit{eROSITA} have not been made, but fact that we find a higher detection rate for \textit{eROSITA} than \textit{LSST} seems surprising, given the former instrument's comparatively low cadence. However, this is due to the fact that the MBHB TDE rate distribution is dominated by high mass MBHBs (see Figs. \ref{fig:dNdlogTDE} and \ref{fig:MBHcontours}), which will produce events which tend to be brighter in the X-ray than the optical (see Fig.\ \ref{fig:bandLums}). The fact that the TDE rate distribution is dominated by lower masses when considering disruptions from single MBHs (see Fig.\ \ref{fig:MBHcontours}) is similarly responsible for \textit{LSST}'s out-performance of \textit{eROSITA}, with events  in this mass range being relatively brighter in the optical. The overall effect of this is that as many as 9.6 per cent of the TDEs detected by \textit{eROSITA} could be from binaries (compared to as many as 1.1 per cent for \textit{LSST}).
\par
Considering methods for actually identifying TDEs in the data from the surveys we have studied is beyond the scope of this work. However, recent work by \citet{deckerfrench18} proposes the search for quiescent Balmer-strong galaxies (particularly the post-starburst subset of these) as a means of identifying TDE hosts with \textit{LSST}, since previous work \citep[e.g.][]{arcavi14,deckerfrench17,graur18} has found such galaxies to be overrepresented among TDE hosts, relative to their scarcity among all galaxies. \citet{deckerfrench18} suggest that 36--75 per cent of TDEs occur in such hosts, which their method can identify with 8 per cent completeness. Applying these results to our own \citep[using the same reasoning they adopt in applying their results to the \textit{LSST} detection rate found by][]{vanvelzen11}, we would expect 81--379 of the disruptions observable yearly by \textit{LSST} to be identifiable in this way. Given that galaxies of this class are likely to have undergone recent mergers \citep[e.g.][]{wild09}, one would expect them to contain MBHBs, and thus that the identification method explored by \citet{deckerfrench18} would favour MBHB induced disruptions. However, \citet{deckerfrench17} suggest that these galaxies are likely the product of mergers with a ratio of progenitor masses of $>1/12$, and that the TDE rate enhancement model of \citet{chen11} would thus not be the main cause for these galaxies having higher TDE rates. In fact, \citet{madigan18} suggest that the formation of an eccentric nuclear disc of stars could be a source of TDE rate enhancement in recently merged galaxies of this type, leading to their over-representation amongst TDE hosts. 
\par
In terms of observational signatures, we considered recurrent TDE flares in detail. Here, \textit{LSST} is likely to perform slightly better (in terms of total number) than \textit{eROSITA}, even if this is only due to the longer survey duration of the former. We expect around 33 galaxies containing MBHBs to disrupt twice over \textit{LSST}'s 10 year survey, compared to only 23 over the four year \textit{eROSITA} run. A similar calculation is done by \citet{weggbode11}, who find 3 MBHBs disrupting twice over 5 years. Using their scaling relation, this would translate to $3\times(10~\text{yr}/5~\text{yr})^2=12$ MBHBs disrupting twice of 10 years. As such, our result is a few times larger than theirs, consistent with the difference between our detection rate calculations. Again, no comparable calculation exists for \textit{eROSITA}.
\par
Although we expect \textit{eROSITA} to detect a smaller total number of recurrent flares from binaries than \textit{LSST}, the galaxies which the former observes as hosting multiple disruptions are actually more likely to contain binaries. We expect only 1.3--2.1 per cent of the MBHB host galaxies inferred from \textit{eROSITA} in this way to be contaminated (i.e.\ actually containing a single MBH). With \textit{LSST}, the level of contamination by single black holes could be between 16.9 and 42.8 per cent, making this sample far less reliable. However, it is quite possible that both samples could be contaminated by galaxies in which a single MBH has captured and disrupted both members of a stellar binary \citep{mandel15}. Such a scenario is likely to lead to fairly rapid disruption of both members, with the time separating the two disruptions tending to be shorter than 150~d \citep{wu18}, thus creating a different light curve to a typical TDE \citep[see][fig.\ 4]{mandel15}. However, a similar light curve could in principle arise from the kind of recurrent flare we consider, in the case of a second disruption happening quickly after the first. One could expect this to be more of a problem for \textit{eROSITA}, since its short survey duration could mean that rapid recurrences form a more significant fraction of its sample.
\par
Given this, it seems safe to say that the observation of three subsequent disruptions in a galaxy should be searched for, in order to provide a more unambiguous sample of galactic centre MBHBs. We expect \textit{LSST} to find around 12 MBHBs in this way, with \textit{eROSITA} finding around 3. As discussed in Section \ref{sec:distinction}, these samples should be totally uncontaminated by single MBHs, with there being almost negligible probability of a single MBH triggering this many disruptions over either survey's lifetime.
\par
Besides recurrent flares, interrupted flares due to perturbation by the secondary MBH in the binary constitute another appealing observational signature. We find that the separation of the MBHBs considered here, although still sub-pc, is likely too large to make such a signature ubiquitous, unless very eccentric binaries are the norm. Other signatures not considered in this work include: i) a shorter duration and steeper fading of the light curve, and ii) stars disrupted by the secondary hole in a binary with $M_1>M_{\rm crit}$. The former has been proposed by \cite{darbha18}, who investigate mpc-separated MBHBs. Deviations from the standard fallback rate should be relatively minor for the wider MBHBs considered here, which typically feed stars into the primary MBH loss cone on relatively wide orbits. The latter cannot be calculated with our current set-up, since \cite{chen11} only derived scaling for stars disrupted by the primary MBH, noticing that, for the fairly unequal mass binaries they considered, the secondary TDE rate is generally lower. We can, however, estimate this contribution using the work of \cite{fragione18}.
\par
\citeauthor{fragione18} predict a TDE rate of $10^{-4}-10^{-3}~\text{yr}^{-1}$ for the secondary, with most disruptions occurring within a 0.5~Myr timescale. We therefore take the upper bound $10^{-3}~\text{yr}^{-1}$ as a fiducial rate for all secondaries, and multiply this by the enhanced phase lifetime given by equation (\ref{eq:tdutycycle}), to give an approximation of the number of TDEs a secondary will disrupt in its lifetime. From this, we estimate the global TDE rate from secondaries to be $(30,80,500)~\text{yr}^{-1}$ for $\gamma=(2,1.75,1.13)$. The increased rate for low $\gamma$ is due to the increased length of the enhanced phase and lack of compensating $\gamma$ dependence in the TDE rate used. Even though this is likely to be an overestimate of the secondary rates, they contribute at most 5\% of total disruptions globally. From Table \ref{tab:resultstable}, we see that around 0.1-0.3\% of all MBHB TDEs are observable -- applying this, we expect observation rates of $\lesssim 1\text{yr}^{-1}$, even with our generous assumptions.
\par
In systems with a sufficiently massive primary that a disruption should not be observed due to direct stellar capture \citep[$M_1 \gtrsim 7\times10^8M_\odot$ would rule out disruption even by a highly spinning black hole, see][]{2012PhRvD..85b4037K}, the observation of any disruptions would act as alternative evidence for the presence of an MBHB -- since in such a case, an observable disruption could only have been caused by a less massive secondary BH \citep{coughlinarmitage18,fragione18}. We calculate only a few such events to occur globally, however: $(1,6,73)~\text{yr}^{-1}$ for $\gamma=(2,1.75,1.13)$. This means an observation rate below $0.2~\text{yr}^{-1}$ even in the best case.
\par
Finally, we identify a number of caveats arising from the assumptions made in our calculation. First, the stellar distribution models used here are normalized to the isothermal sphere outside the influence radius of the binary, resulting in relatively dense stellar profiles. Although the total number of TDEs might not be very sensitive to the assumption of shallower density profiles, the number of detected recurrent flares certainly is. Nonetheless, we note that the isothermal sphere is a reasonable approximation to the density profile of the Milky Way, and stellar cusps are expected to be relatively common in Milky Way-like galaxies which are the preferred hosts of MBHB TDEs. Second, we focused on the cusp-erosion phase only in determining the MBHB TDEs. As shown in Appendix \ref{app:nonenhancedbinaries} this might or might not be a good proxy for the total TDE rate depending, in particular, on the slope profile of the stellar cusp $\gamma$. TDEs from the subsequent hard binary phase might dominate the overall rate for shallow density profiles ($\gamma \lesssim 1.5$) -- so the rates reported here are a safe lower limit and can be higher by up to a factor of a few.
Third, we assume that, at some point in the TDE evolution, every component of the emitted spectrum is observable. This assumption is clearly falsified if the nature of the observed emission depends on the viewing angle to the observer \citep{2018ApJ...859L..20D}. In this case, all the numbers derived in this work should be roughly divided by a factor of two. Fourth, we did not try to evaluate the statistical significance of TDE flare detection. This is expected in particular to have a strong impact on the fidelity of {\it eROSITA} candidates. The eRASS survey strategy implies only $\sim 10$ combined points in the lightcurve of each target. Identifying recurrent flares at high statistical significance from a handful of points in the lightcurve might be very challenging. Dedicated follow-up (with other instruments) after the first outburst is identified can alleviate this issue. Last, although the fact that our MBH population model yields numbers of regular TDEs consistent with previous investigation is a good sanity check, the number of detected MBHB TDE flares critically depends on the unknown cosmic population of unequal mass, sub-pc MBHBs. If stalling of unequal MBHBs is common, then the number of binary induced TDEs will be greatly reduced; still, a sizeable number of events are expected from systems with $q>0.1$, as shown in Fig.\ \ref{fig:dNdlogTDE}.
\par
Mindful of those caveats, we have demonstrated that sub-pc MBHB induced TDEs should lurk in large numbers in the wealth of TDE candidates that future optical and X-ray surveys will identify. Our ability to separate these signals from standard TDEs might be the key to unveiling the unexplored cosmic population of sub-pc MBHBs in the mass range $10^5\msun<M<10^7\msun$. The unique appeal of this prospect resides in two key facts. First, due to their small mass, it might be extremely hard to discover these systems at cosmological distances via other electro-magnetic observations. Second, this is the mass range relevant to future gravitational wave observations with {\it LISA} \citep{2017arXiv170200786A}. The identification of MBHB TDEs will therefore critically inform the future of low frequency gravitational wave astronomy.

\section*{acknowledgements}
We thank X.\ Chen, E.M.\ Rossi and G.P.\ Smith for useful discussions. We also thank N.C.\ Stone, T.\ Wevers and the anonymous referee for their comments on the manuscript. ST also acknowledges D.\ Muthukrishna for highlighting the current situation regarding \textit{LSST}'s cadence. AS is supported by the Royal Society.

\bibliographystyle{mnras}
\bibliography{bib.bib}

\appendix
\section{Disruptions from MBHBs Outside of the Enhanced Phase}
\label{app:nonenhancedbinaries}
      In this paper, we focus on the phase of cusp erosion following the MBHB merger \citep[e.g.][]{2007ApJ...656..879M,sesana08}, which is generally insufficient to lead the MBHB to final coalescence. Hardening will further continue due to scattering of unbound stars within the binary loss cone. Those dynamical interactions will continue to produce MBHB TDEs, albeit at a lower rate, so that identification via multiple flares becomes highly unlikely. Conversely, the binary is more compact in this stage, enhancing the chance of identification via flare interruptions.
      
      Although we have omitted these events from our study, they can still contribute significantly or even dominate the overall TDE rate from MBHBs. For a simple estimate, we use equations 6 and 7 in \cite{sesanakhan15} to estimate the typical lifetime of a binary in the hard phase, which can span three orders of magnitude from 1~Myr to 1~Gyr, depending on $M_1$, $q$ and, most importantly, $\gamma$. This is then multiplied by the single MBH TDE rate given by equation  (\ref{eq:ntdsinglee}), to get the number of TDEs, ${\cal N}_h$ occurring in the hard phase as a function of the system parameters. This is justified by theoretical work that finds that hard binaries disrupt stars at a lower rate \citep{2008ApJ...676...54C}, possibly comparable to or slightly higher than the typical rate for single MBHs \citep{coughlin18,darbha18}.

\begin{figure}
\includegraphics[width=\linewidth]{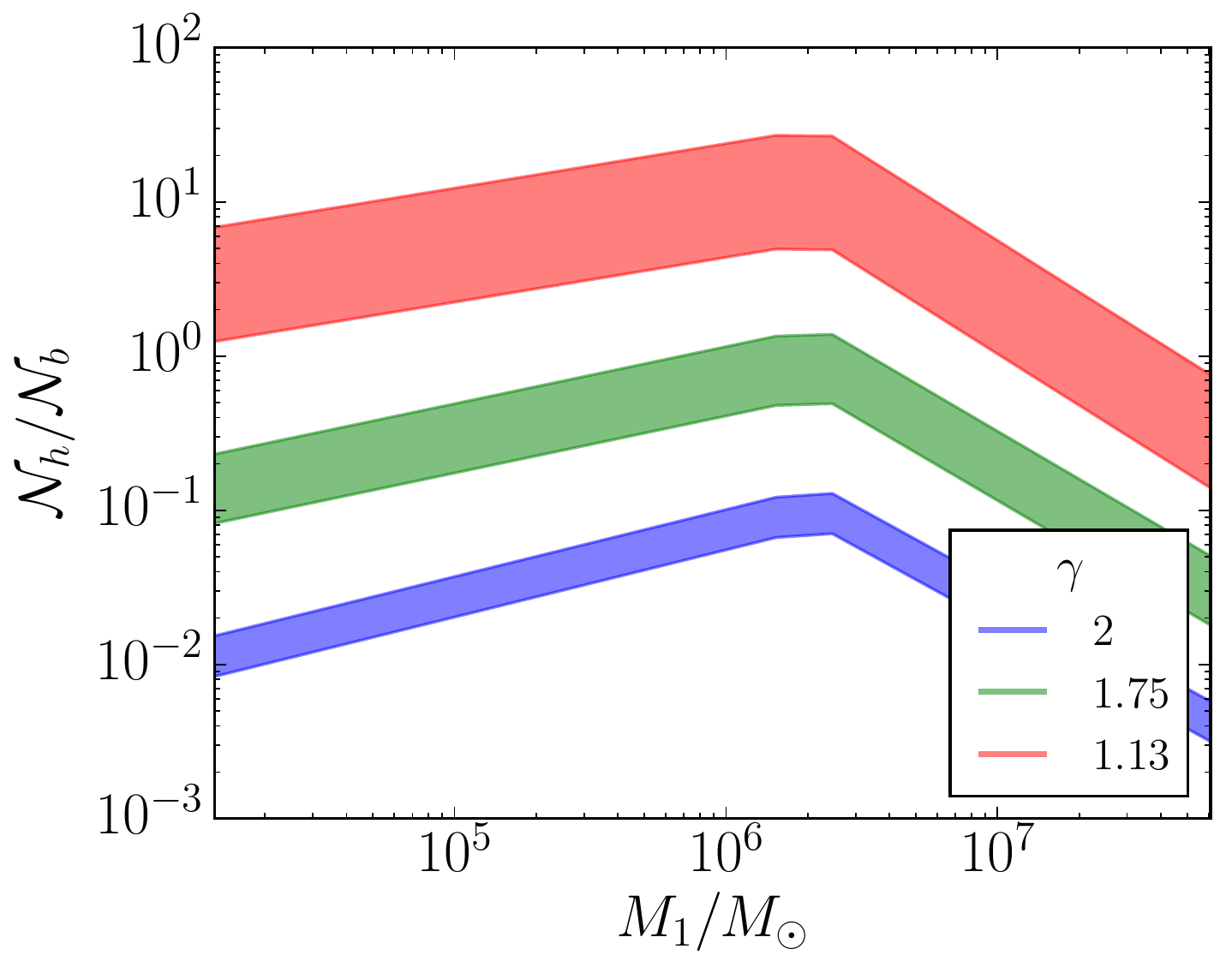}
\caption{Ratio of TDEs that occur outside of a binary's enhanced phase to those that occur during the enhanced phase. The vertical spread comes from the range of q-values used in this paper (with the top of the band corresponding to $q=0.003$, and the bottom to $q=1$).}
\label{fig:fracs}
\end{figure}

        The number of TDEs so obtained, is then compared to those in the enhanced, cusp erosion phase, ${\cal N}_b$ given by equation \eqref{eq:ntd}. The ratio ${\cal N}_h/{\cal N}_b$ is plotted in Figure \ref{fig:fracs} as a function of $M_1$ and for different $\gamma$. For $\gamma=1.75$ and above, the ratio is below 1 for all, or essentially all, combinations of $M_1$ and $q$, and the TDE rates are in fact dominated by the cusp erosion phase. However, for $\gamma=1.13$, the ratio is often above 1, and can be as high as 10. The transition between the two regimes likely happens somewhere around $\gamma=1.5$. Therefore, strictly speaking, using equation (\ref{eq:ntd}) to describe the total number of TDEs by an MBHB is only justified for systems residing in cuspy galaxies. For those hosted in galaxies with shallower cusps, equation (\ref{eq:ntd}) is only a robust lower limit to the total TDE rate, which can be a factor of a few higher.

\section{Impact of Possible Emission Scenarios}
\label{app:emissionmodels}
In section \ref{sec:emission}, we construct a purely empirical model for TDE emission -- aiming for agreement with observed luminosities, rather than an accurate physical description. Many physically motivated emission models do exist in the literature, though, so it is interesting to consider the impact they could have on our results.

Since Eddington limited accretion is not typically sufficient to describe the observed optical emission, one popular explanation which is frequently invoked is the presence of winds caused by super-Eddington accretion \citep[e.g.][]{strubbe09,strubbe11}. \citet{2018MNRAS.478.3016W} predict that black holes with $M_\text{BH} < 10^7M_\odot$ can accrete at a super-Eddington rate for months to years. This means that a large fraction of the MBHs producing TDEs (more so in the case of lone MBHs, where the majority will be in this mass range -- see Fig.~\ref{fig:massfuncs}) could experience a substantial phase of super-Eddington accretion. The $g$-band and bolometric luminosities predicted by  \citet{strubbe09} are broadly consistent with our phenomenological model (compare their fig.~4 to our Fig.~\ref{fig:bandLums}), with the former falling mostly between $10^{42}$ and $10^{43}$~erg\,s$^{-1}$ for all masses and the latter in between $10^{43}$ and $10^{45}$~erg\,s$^{-1}$. There is, however a stronger mass dependence predicted by the \citeauthor{strubbe09} model, which is additionally sensitive to how close the star approaches the black hole. This is unlikely to affect the total detection rates, as the predicted variation with mass falls within a range of $L_g$ roughly consistent with our model. In fact, the \textit{LSST} detection rate of 6000~yr$^{-1}$ predicted by \citeauthor{strubbe09} is comparable to ours, albeit slightly higher since they require observability in the $g$-band only -- compare also their fig.~13 and our Fig.~\ref{fig:obsmassfns}.

Alternative models include that of \citet{coughlin14}, who predict that excess energy is not carried away by winds, but instead contributes to the inflation of a gaseous envelope, from which energy is eventually released via jets piercing through the poles. The effective surface temperature of this envelope is predicted to fall approximately between 35000~K and 50000~K for MBH masses between $10^5$ and $10^6M_\odot$ (it would only be a factor of a few higher extending to $10^7M_\odot$). If the envelope emits as a black body, this would correspond to emission peaking around a $\text{few}\times10^{15}$~Hz -- roughly consistent with the optical black body component of our phenomenological model. This is not surprising, since the temperature range predicted by \citet{coughlin14} aligns with the black body temperatures obtained observationally by \citet{wevers17}. Similarly, the circularisation shock model of \citet{piran15} predicts effective black body temperatures giving rise to optical luminosities consistent with our phenomenological model.

\bsp	
\label{lastpage}
\end{document}